\documentclass{ws-rv9x6}
\usepackage{subfigure}   
\usepackage{ws-rv-thm}   
\usepackage{ws-rv-van}   
\makeindex

\usepackage{ulem}

\begin{document}

\chapter[The Survival of Ernst Ising and the Struggle to Solve His Model]{The Survival of Ernst Ising and\\ the Struggle to Solve His Model\label{ra_ch1}}

\author[R. Folk]{Reinhard Folk\footnote{r.folk@liwest.at}}

\address{Johannes Kepler University\\Institute for Theoretical Physics\\
Altenbergerstr. 69, 4040 Linz, Austria}

\begin{abstract}
The life of Ernst Ising and the steps to solving the model named after him are reported in parallel. Wilhelm Lenz suggested his student Ernst Ising to explain the existence of ferromagnetism on the basis of his publication in 1920. The result, published in 1925 was disappointing, because only the one dimensional case could be solved with a negative result about the absence of ferromagnetism. Wolfgang Pauli who was an assistant of Lenz in Hamburg published in the same year his ‘nonclassical ambiguity’, later identified as the spin of the electron, and the exclusion principle. He was the first - at the Solvay Conference in 1930 - to present the Hamiltonian of the Ising model as we know it today.

Meanwhile Ising had left university research and due to the political situation in 1938 had to leave Germany and fled to Luxemburg. This went in hand with damaging the network of researchers dealing with the problem of ferromagnetism and more generally with phase transitions and statistical physics. In 1944, the year when Luxemburg was liberated by the American troops and Ising and his family was rescued, Lars Onsager presented a solution of the two-dimensional case. In 1952 Chen-Ning Yang solved the problem of Ising’s thesis in two dimensions; one year later Ising became the US citizenship. The following development showed, that the model turned out to be a highway to modern physics concepts applicable also in other fields, although the final exact solution in three dimensions has not yet been reached.
\end{abstract}


\body

\tableofcontents

\newpage

\section{Introduction}  

{\it History is the most fundamental science, for there is no human knowledge which cannot lose its scientific character when men forget the conditions under which it originated, the questions which it answered, and the functions it was created to serve.}[Benjamin Farrington \cite{farrington} cited by Erwin Schr\"odinger \cite{schroedinger1}]\vspace{0.2cm}

The quest to understand the emergence of {\sc magnetism in solids} after the Great war is entangled to the worldwide changes  in the social and political environment of the involved scientists. 
It started from a simple idea for three-dimensional magnets of a German university professor in 1920 after the First World War and ended in 1952 in the United States with the solution for the two-dimensional case by a university professor born 1922 in China. 
No solution for the three-dimensional case is known so far.
But the model formulated by Ernst Ising has become a  physical standard model reaching out in other physical fields and other disciplines different from physics like chemistry, biology, medicine, ecology, computer sciences, economy, social sciences,  linguistic and much more. In physics it lead to merging ideas of solid state physics with ideas of elementary particle physics and is even fruitful for modelling our understanding of space and time. Nevertheless one find's opinions like this: \vspace{0.2cm}

{\it This model was suggested to Ising by his thesis adviser, Lenz. Ising
solved the one-dimensional model, ..., and on the basis of the fact
that the one-dimensional model had no phase transition, he asserted
that there was no phase transition in any dimension. As we shall
see, this is false. It is ironic that on the basis of an elementary
calculation and erroneous conclusion, Ising’s name has become among
the most commonly mentioned in the theoretical physics literature.
But history has had its revenge. Ising's name, which is correctly
pronounced “E-zing,” is almost universally mispronounced “I-zing.”}
[Barry Simon \cite{simon}]\vspace{0.2cm}

One goal of this publication is to make understandable why Ernst Ising could only solve the one dimensional model and why this was at his time not just a student exercise as it is nowadays.  On the other hand  the development of physics is followed and the steps made by now well known scientists to reach a solution are shown. To solve the model in higher dimensions Alexander M. Polyakov\cite{polyakov} named two miracles, which were necessary to happen, a small one - named duality - and a large one - the Onsager solution (impossible to predict the mathematical apparatus necessary). From this it became clear that Ising's task was a mission impossible and it remains so far. 

In 2011 McCoy said about {\it The Romance of the Ising Model} \cite{McCoy1}: "Fortunately for romance, there are many mysteries of the Ising model which are far from being understood. The romantic in me says that, even when these mysteries have been understood, the understanding of the mysteries will generate new mysteries and the romance of the Ising model will be everlasting."

And in the introduction to his talk {\it The Once and Future Ising Model} given at the Simons Center for Geometry and Physics on December 15, 2015 he cites the mathematician Charles Dodgson (better known from literature as Lewis Carroll) when he explained that his talk is not about results but problems with Ising's model: \vspace{0.5cm}

\parbox{0.26\textwidth}{
\includegraphics[width=0.3\textwidth]{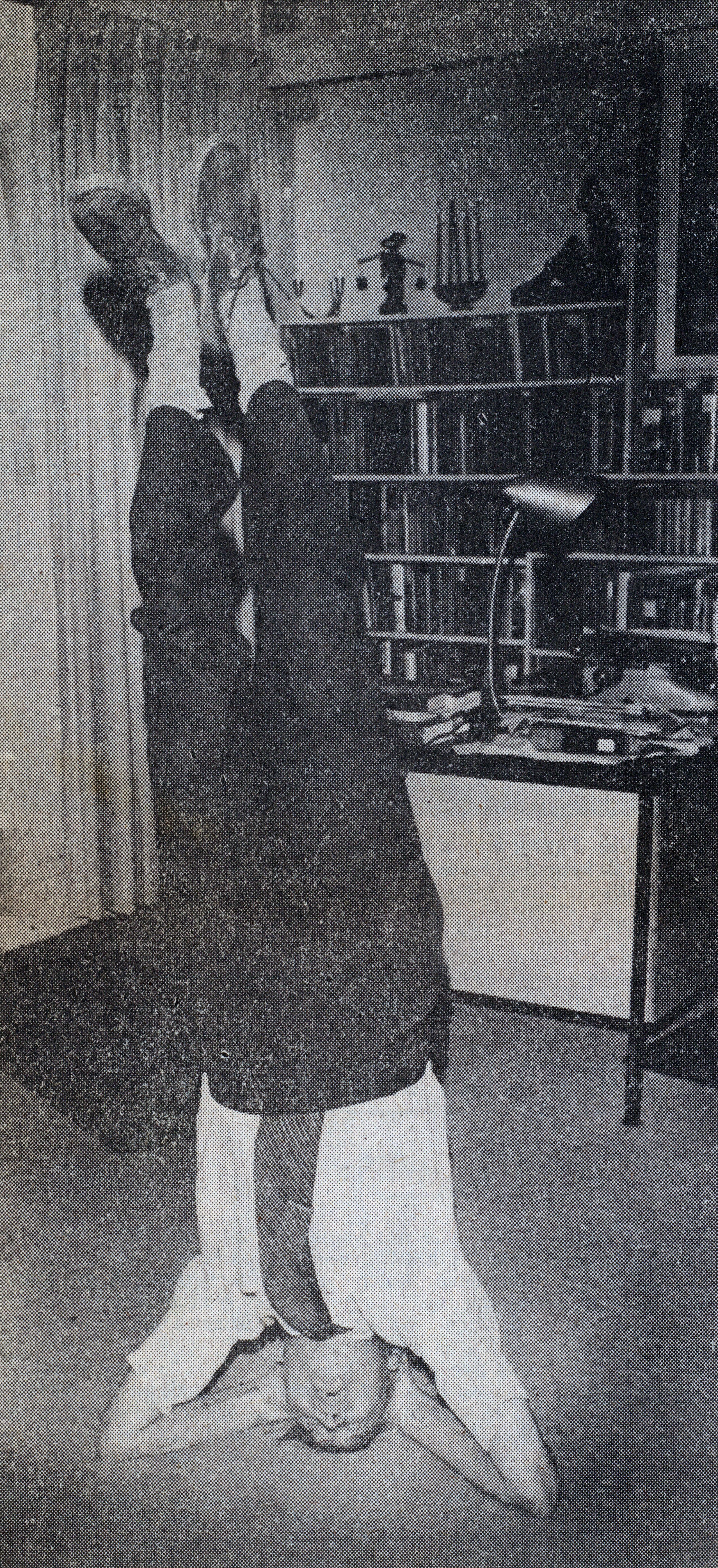}\\
Ising 1976\\ {\tiny \copyright Peoria Journal Star}}
\hspace{0.01\textwidth}
\parbox{0.73\textwidth}{
\begin{verse}
``You are old, Father William," the young man said,\\
``And your hair has become very white;\\
And yet you incessantly stand on your head –\\
Do you think, at your age, it is right?” \\
``In my youth," Father William replied to his son,\\
``I feared it might injure the brain;\\
But, now that I’m perfectly sure I have none,\\
Why, I do it again and again."\vspace{0.3cm}\\ 
(Lewis Carroll 1865) \end{verse}}

\section{Another Introduction}

{\it History uncovers the strange stew of concepts that were necessary for the development of physics.}[M. Stanley \cite{mstanley}]\vspace{0.2cm}

In the year 1922 when Chen Ning Yang - who solved and published the solution of Ising's problem in two dimensions - was born, Wilhelm Lenz asked his doctoral candidate Ernst Ising at Hamburg University to start with his thesis. Ising had to formulate a model which should replace [supersede] the Curie-Weiss theory \cite{weiss1} explaining the existence of a ferromagnetic phase in solids. It was already known \cite{Bohr,vanLeeuwen} that ferromagnetism is a quantum effect but when Ising started his thesis the formalism of quantum mechanics was just in the revolutionary change from the old pseudo-classical version of Bohr-Sommerfeld to its new abstract version of Pauli, Heisenberg and Schr\"odinger.   

Two years earlier another young man named Lars Onsager was admitted to  start his studies at  the Technical University of Trondheim. He came in 1928 to the United States where he later in 1944 published the exact solution of the Ising model in two dimensions and zero magnetic field.  


In 1938 Ising had to flee from the Nazis to Luxemburg and emigrated in 1947 to the States. Yang went to the States in 1946 not returning to the communist China. Similar fates can be told from some of those who filled the gap between the formulation of the problem and its solution, like for Bethe\cite{wolff1993}
``You will probably not know that my mother is Jewish: According to the civil service law, I am 'not of Aryan descent' and consequently not worthy of being a civil servant in the German Reich. [...] So I have to face the consequences and try to find somewhere abroad.'' He left Germany in 1935 as Peierls in 1936 and worked with him in Manchester. Others got in touch with the net which was thrown from Europe to the United States in that time like Montroll, Wannier or Onsager. Again others who stayed in Europe  tried to survive the Second world War and keep the contact they have established before, like Kramers. 

In 1930 physics was very different from how it looks today. Only two forces (gravity and electromagnetism), and three particles (the proton, the electron and the photon) were known at that time. In order to uncover these concepts we have to look at the past but avoid to look at it through the lens of the present. But we also have to uncover which concepts known in the past have been used by the scientists,  to understand how they tried to solve the open problems and to learn from their errors.

Marc Kac who was granted a scholarship to the United States after finishing his Ph.D. in Lviv gave a combinatorial solution to the two-dimensional Ising model remarked in his autobiography\cite{kac}: 

``To explain phase transitions from `first principles' on the basis of statistical mechanics is still a largely unsolved problem. It continues to fascinate physicists and since it can be stated in purely mathematical terms it also attracted the attention of mathematicians.

Because of the inherent complexity of matter one is compelled to work with simplified models which, one hopes, capture the qualitative and even some quantitative features of their counterparts in nature.  

The most celebrated class of models proposed for the purpose of understanding magnetic phase transitions is the Ising models. Introduced in 1925 by Ernst Ising, they are still vigorously studied today and the literature on them is staggering."

And quite recently Ising's one-dimensional model is reconsidered and a solution by mathematical induction has been suggested \cite{wang} because: ``The one-dimensional (1D) Ising model is of fundamental importance in an introductory course on statistical mechanics, because it is connected to many interesting physical concepts...  Despite the absence of a genuine phase transition, the 1D Ising model still plays a central role in the comprehension of many principles, phenomena, and numerical methods in statistical physics."


In the following alternately the lifetime of Ising and the development in physics related to his thesis is described.



\section{Starting Ising's academic education}

 \parbox{0.30\textwidth}{
\includegraphics[height=5.3cm]{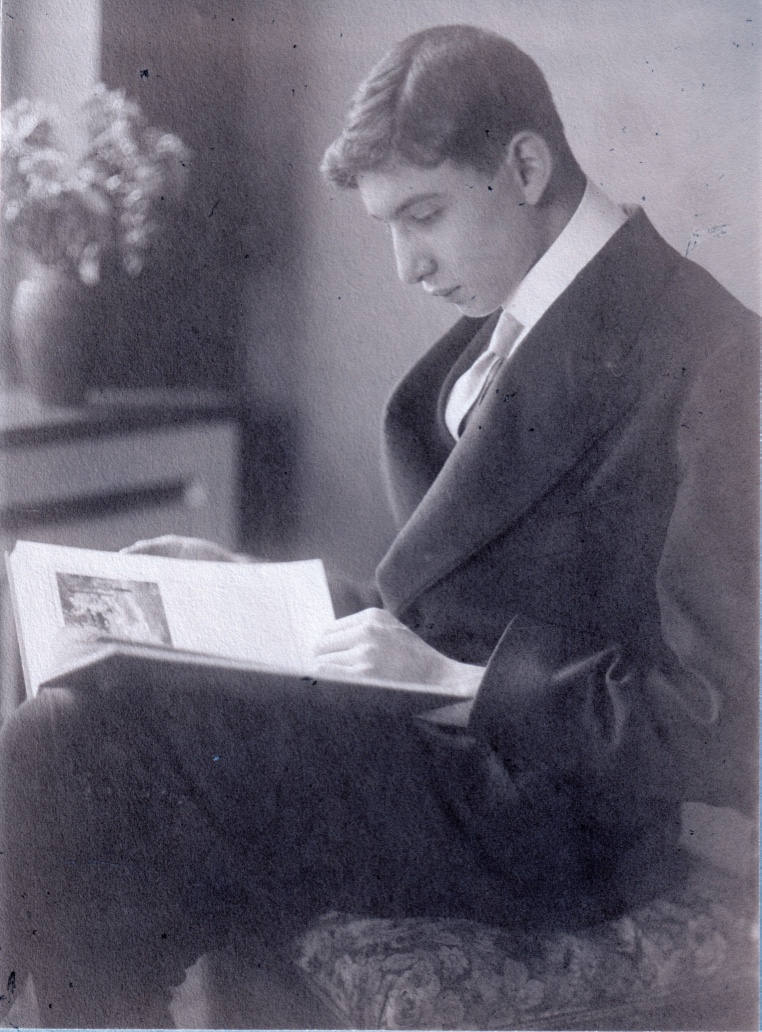}\\
Ernst Ising in Bochum\\ {\tiny\copyright Tom Ising}}
\hspace{0.03\textwidth}
\parbox{0.65\textwidth}{
 \begin{itemize}
 \item[1900] born in Cologne as the son of Gustav and Thekla Ising, a Jewish merchant family 
\item[1904] living in Bochum, Goethestra{\ss}e 18
\item[1918] Matura (general qualification  for university entrance), military training
\item[1919]  begin of the studies of  physics and mathematics at the University of G\"ottingen, one semester interruption, then studies of astronomy at the University of Bonn 
 \item[1921] April 21 starting studies at the University of Hamburg
 \item[1924] December finishing the PhD thesis at the University of Hamburg
  \end{itemize}} \vspace{0.5cm}

Ernst Ising was born on May 10, 1900 to the Jewish family of  Gustav Ising and his wife Thekla L\"owe. His father was a merchant in Cologne [Heumarkt 64/66]. Ernst completed his schooltime in Bochum and after the military education started his study in mathematics and physics  in the summer semester 1919 at the University of G\"ottingen\cite{goettingen}.
``Under Klein, Hilbert, Minkowski, Landau and Runge, together with the astronomer Karl Schwarzschild and the physicists Ludwig Prandtl, Peter Debye and Emil Wiechert, there formed the creative atmosphere, described vividly by Max Born, Harald Bohr and Richard Courant, which attracted scientists and students from all over the world and made G\"ottingen the Mecca of mathematics.'' 

After a break of one semester may be a so called `Zwischensemester'  \cite{geppert}  he proceeded with mathematics at the university of Bonn \cite{bonn} (22.04.1920) where he also studied astronomy\cite{curriculum}. Two semesters later he moved \cite{reich} on the 21st  April 1921 to the new University of Hamburg founded in 2019.  There  he turned toward theoretical physics on recommendation of professor Wilhelm Lenz (1888 - 1957) and started end of 1922 with his thesis under Lenz'  instruction.

Meyenn \cite{meyenn1}  describes the situation of physics in Germany: 
``Through a skilful science policy based on the principle of self-government, and the strong mathematical orientation of the exact natural sciences, which originated especially from G\"ottingen, the theoretical physicists of the Weimar Republic were able, despite the general misery after the First World War, extremely successfully to contribute to the clarification of the atomic riddle. This glittering epoch of theoretical physics, which culminated with the establishment of quantum mechanics, has therefore been described as the golden twenties.

A development that could neither be stopped by the war nor by the economic plight of the post-war years, then came to a standstill within a few years due to the anti-science policy of the National Socialists. Arnold Sommerfeld, a physicist who had been at the center of this development through his research and teaching, wrote resignedly to Einstein in September 1937: {\it The future looks bleak for German physics. I must console myself with having experienced her golden age 1905 - 1930.}"

One reason for the success might have been the fact that Theoretical Physics was `cheap' compared to the upcoming large research facilities in the experimental sector.

\subsection{Hamburg University and the Institute for Theoretical Physics}

Despite the great misery in Germany after the First World War, a new university was democratically founded in Hamburg\cite{lenzen2019}: ``1919, the year the University was founded, saw a longstanding debate come to an end in Imperial Germany about whether a university was of any use to the port and trading city of Hamburg. Before the German Revolution of 1918/19, the majority of the city's decision-makers had persistently rejected the idea. Yet on  March 28, 1919, the time was ripe. A social democrat majority in the Hamburg parliament succeeded in founding the University with a sudden shift in their argument: It was no longer about the usefulness of a university for merchants and shipping companies but whether university education could dismantle class differences. This was precisely what the protagonists of the University's foundation had in mind: by opening a university for all, it would be possible to guarantee education and, as a consequence, socioeconomic progress for all, which would ultimately end class differences.'' 

One year after the foundation an Institute for Theoretical Physics was created and Wilhelm Lenz was appointed 1921 as Ordinarius Professor.  He had graduated in Munich under Arnold Sommerfeld, became his assistant and completed there his Habilitation. In the following he got the position of an extra-ordinarius professor in Rostock from where he moved to Hamburg as director of the Institute of Theoretical Physics. There Lenz established the reputation of his research with a small group of doctorands together with  Wolfgang Pauli, who also came from Sommerfeld's group in Munich as research assistant  for habilitation in Hamburg.

In general the university of Hamburg and especially the physics department became internationally recognized. Rabi who was one year at Hamburg University wrote\cite{rabi}: ``When I was at Hamburg University, it was one of the leading centers of physics in the world. There was a close collaboration between Stern and Pauli, between experiment and theory. For example, Stern's questions were important in Pauli's theory of magnetism of free electrons in metals. Conversely, Pauli's theoretical researches were important influences in Stern’s thinking. Further, Stern's and Pauli's presence attracted many illustrious visitors to Hamburg. Bohr and Ehrenfest were frequent visitors.''

\subsection{At Lenz' institute}
The conditions for doctoral candidates have been described\cite{lenz60} as very good: 
``Joh[annes Hans Daniel] Jensen [(1907 - 1973)], at the time a student at the seminary, told me that Lenz never stifled his time when it came to helping his disciples; you could sit with him until late into the night and consult with him. He did not settle for anything until the basic features of a problem were clearly and simply worked out."

When Ernst Ising joined the group of Lenz Werner Schr\"oder had already finished his thesis\cite{schroeder22} and also Lucy Mensing joined the group. Access for women to universities was made easier in the 1920s, but was tightened again in the 1930s under the National Socialist dictatorship. She worked on a quantum mechanical problem\cite{mensing25} and was mainly supervised by Wolfgang Pauli, with whom she later worked on the new quantum mechanics\cite{muenster}.  This time together at the institute with Ernst led to a lifelong friendship. 
Ising himself describes his situation\cite{brush} at the institute as follows: ``At the time I wrote my doctor thesis Stern and Gerlach were working in the same institute on their famous experiment on space quantization\cite{frankfurt}. The ideas we had at that time were that atoms or molecules of magnets had magnetic dipoles and that these dipoles had a limited number of orientations. We assumed that the field of these dipoles would die down fast enough so that only interactions should be taken into account, at least in the first order \ldots I discussed the result of my paper widely with Professor Lenz and with Dr. Wolfgang Pauli, who at that time was teaching in Hamburg. There was some disappointment that the linear model did not show the expected ferromagnetic properties." Beside those discussions Ising could also communicate with Mensing on the quantum mechanical background of his model.

Regarding Pauli Meyenn \cite{paulimeyen} commented Pauli's aversion to solid-state physics but explained:  ``This is all the more remarkable given that he himself gave the first impetus to the development of the quantum theory of the solid with his theory of paramagnetism  and also suggested a solid-state theme for Peierl's dissertation [{\it Zur kinetischen Theorie der W\"armeleitung in Kristallen (On the kinetic theory of heat conduction in crystals)}]. Later, too, Pauli included solid state physics in his lecture (in the summer semester of 1938) and also took part in such events with interest.  What Pauli fundamentally rejected was not solid-state physics as such, but the description of physical effects by approximation methods or as a consequence of seemingly uncontrollable circumstances."

\section{The way to Ising's thesis}

{\it Science progresses discovery by discovery.}[St. Thurner, W. Liu, P. Klimek, S. A. Cheong\cite{thurner1}]\vspace{0.02cm} 

At the beginning of the 20th century thermodynamic theories were developed to give a practical quantitative description of the change of phases in condensed matter with temperature. The first was the van der Waals theory with its equation of state for gaseous and liquid phases in a fluid. The key of finding an equation of state was the introduction of an internal pressure resulting from the interaction between atoms or molecules. As a result two phases of different density appear, which at the so-called critical point at a certain pressure and temperature could continuously change into each other by going from the high temperature gas phase to the low temperature liquid phase.

A quite similar phenomenon was observed in solids which could be magnetized by an external field and stay magnetized when the field is turned off.  Increasing the temperature, the magnetization became weaker and vanished at a certain temperature, the Curie temperature. This was called ferromagnetism in contrast to para- and diamagnetism where the solid system remained unmagnetized when the external field was turned off.

It was Pierre Weiss  who introduced in 1908 the concept of the internal molecular field\cite{weiss1,weiss2} in which the elementary magnets find their ordering and later he related this field to the internal pressure in fluids. This brought already at the beginning of studying phase transitions a kind of {\it universality} to the phenomena\cite{berche2009}.

\subsection{Kirwan's hypothesis - a lost historical root}

The Curie Weiss theory was a big advance because of its eminent practical use to explain the experimental findings in ferromagnets like the existence of a phase transition, the change in the specific heat when changing the phase, the qualitative increase of the magnetization in the ferromagnetic phase, and  the behavior of the susceptibility. However its central assumption of an internal magnetic field resulting from the arrangement of rotatable elementary magnets and their interactions was unclear and criticizable. Moreover quantitative deviation especially near the Curie point remained.

A report\cite{bulletin} of a committee of the U.S. National Research Council issued by the National Academy of Sciences, Washington on {\it Theories of Magnetism} was published at the time when Ising started with his thesis. A translation\cite{wuerschmidt} was published in 1925 where the authors reworked the copy of the translation and made a number of improvements and additions. One of the authors,
E. M. Terry, explains the nature of the molecular field: ``Just as in the case of a gas, to account for the transition to the liquid state, there must be added to the external pressure an internal one due to
the mutual attractions between the molecules, so in the case of a ferromagnetic substance, as it is cooled
in a magnetic field from a temperature which has rendered it paramagnetic, the transition to the ferromagnetic state is explained by assuming that,  due to the overlapping of the fields of the individual molecules, there comes into existence an internal or molecular field, which added to the external field, accounts for the very large intensity characteristic of  this state.''

However the nature of the internal (molecular) field remained unclear and its strength turned out to be too small in order to lead to the observed Curie temperatures. Terry\cite{bulletin} wrote: ``It thus appears that the molecular field can have neither a magnetic nor an electromagnetic origin and must therefore be of a nature different from the ordinary magnetic fields with which we are familiar.'' Moreover no model in the context of quantum mechanics was available.  J. Kunz\cite{bulletin} discusses different values of the magneton within the Bohr atom and his comment "No theory explains so far why the electron, moving on a circular or elliptic orbit, about a nucleus, does not lose the energy by radiation, or why the orbit remains stationary or free from radiation'' shows the difficulties of applying classical models in quantum mechanics. 

Thus also Ising started his thesis directly with the statement that ferromagnetism is a sofar unexplained phenomenon and opens his historic time line of the different ideas on ferromagnetism with Richard Kirwan (1733 - 1812). 
The first paragraph of Kirwan's paper\cite{kirwan} starts with a statement which may be considered as a motto to Ising's model:

{\it There are two ways of explaining a natural phenomenon; the first, is by discovering the conditions and circumstances of its production and the laws by which its action is governed; the second, is by shewing its analogy, similarity or coincidence with some general facts with whose laws and existence we are already acquainted; this last mode is by far the most perfect and satisfactory. In the first sense of the word electricity and magnetism have been in some measure explained, but in the last sense neither; the primary cause of magnetism in particular has hitherto been supposed to iron alone, or its ores, and to stand unconnected with all other natural phenomena.}

Kirwan an Irish scientists working in the fields of meteorology, geology, chemistry, physics and philology was Fellow of the Royal Society, supporter of the Copley medal and from 1799 until his death president of the Royal Irish Academy. His publication {\it Thoughts on Magnetism}, which Ising cites, appeared 1797 in the Transactions of the Royal Irish Academy\cite{kirwan} and was later translated into German. However the editor remarks:
``I give these ideas, just a little abbreviated as I find them in English, since I believe right away that they are more witty than correct.". This judgement is relativized by the editor at the end of the article and he points to Kirwan's idea to relate the magnetic force to the force of crystallization "yet their {\it direction} in all its varieties being exactly the same". 

One might speculate how Ising came to recognize Kirwan's paper. He may have found it already in Felix Auerbach's {\it Geschichtstafeln der Physik (Historic Tables of Physics)} from 1910. Auerbach collected important progresses in physics ordered in time from -650 to 1900. In this list he cites in the year 1797 for the topic: ``Richtungshypothese f\"ur den Magnetismus" [Directional hypothesis for magnetism] and mentioned Kirwan.   Auerbach was already in 1895  the author of the section on magnetism in the  {\it Handbuch der Physik}\cite{auerhand} [second ed. 1908]. There on page 31 [48] he explained the history of the directional hypothesis starting with Kirwan and further developed by Ohm\cite{ohm} and Weber\cite{weber}. It assumes that there are already elementary magnets but randomly oriented so that the whole magnet is unmagnetized. By orientation in one common direction the solid becomes magnetized.

 
In this way Ising connected the old ideas of Kirwan directly to his model based on the modern development made by Lenz and Schottky.  

\subsection{Lenz's idea of directed magnetons}

In order to explain the ferromagnetic phase in solids Wilhelm Lenz suggested\cite{lenz1920} in 1920 that the elementary magnets in solids are not free rotatable  but are directed in certain directions defined by the geometry of the solid. In order to stay simple he assumed that two directions are allowed, namely the original  direction and its reverse. Thus elementary magnets in the solid have only the possibility to be parallel or antiparallel to each other. For such a two state situation he calculated the susceptibility and reproduced Curie's law.

For ferromagnetic solids he then assumes that there is an energy difference between parallel oriented  and antiparallel oriented neighboring magnets. In consequence an overall ferromagnetic phase may be explained. The energy difference should result from a nonmagnetic interaction between the elementary magnets. 

Lenz was extraordinarius professor at the University Rostock when he published the paper. In December 1920 
the procedure for filling the new  position of a full professor of Theoretical Physics started for which Lenz applied. Einstein was asked about the candidates and wrote in a letter\cite{reich} to Edgar Meyer: ``I estimate Lenz similar to Debye and especially consider his [Lenz's] last, for the time being only very incomplete publicized works on band spectra and magnetism extremely important.'' Finally Wilhelm Lenz got the position and started his job in Hamburg 1922 by organizing a so called {\it Vortragsseminar} (lecturing seminar) where also notable external guests participated.

\subsection{Stern-Gerlach experiment and spatial quantization}

An open question in the Bohr-Sommerfeld quantum mechanics was the quantization of the orientation of the angular momentum and the orientation of  the corresponding magnetic momentum in an external field. In order to solve this problem Otto Stern 1919 suggested an experiment which was  performed in  1922 together with Walther Gerlach at the University of Frankfurt\cite{sauer}: ``At the time when the SGE [Stern Gerlach experiment] was performed, the physics community did not understand why and how the internal magnetic moment (i.e. angular momentum) of each atom {\it collapses} in the SG-apparatus into well-defined angular orientations with respect to the direction of the outer magnetic field. This clearly contradicted classical physics where a Larmor precession of the magnetic moments was expected. For most physicists this was a {\it miraculous interaction} between moving atoms and the SG apparatus."
The outcome of the experiment was interpreted to show spatial or directional quantization but in fact it an additional property of the electron later recognized as the spin of the electron was observed. Pauli sent a post card on February 17, 1922 to Gerlach with the remark\cite{pauliGerlach}: ``This should convert even the nonbeliever [in spatial quantization] Stern."

\subsection{Schottky's synchronous electrons \label{schottky}}

On September 19, 1922 Walter Schottky came from the University of W\"urzburg for the winter semester 1922/23 to the University of Hamburg as a salaried assistant at the Physical State Laboratory. At the Verhandlungen der Gesellschaft Deutscher Naturforscher und Arzte (Meeting of the Society of German Natural Scientists and Physiciens) in Leipzig he presented his results\cite{schottky,schottky1922} published in 1922.  

He corroborated Lenz's idea by suggesting a quantum mechanical model for the energy difference of parallel and antiparallel elementary magnets.  This paper is nowadays known  for a phenomenon in the specific heat he calculated for the two state system presented in his paper named {\it Schottky-anomaly}. The specific heat as a function of temperature shows a peak at low temperature  before it goes to zero instead of going to zero continuously. 

Schottky considered neighboring electrons circulating like planets around their atomic nucleus and asked the question: how have two neighboring electrons to behave in the Bohr Sommerfeld picture in order to minimize their Coulomb energy? He came to the conclusion: ``In the field of magnetism, the introduced roto-active interactions of the atoms seem to provide the previously unknown type of directional force, which produces the self-ordering of the magnetic atom axes and thereby becomes the cause of the spontaneous magnetization and the ferromagnetic behavior of certain substances." In order to show this he calculated the  Coulomb energy of the two  electron circling on their atomic path (`planetary' circles) around the nucleus of the atoms and showed that a minimum is reached, when the electrons rotate in the same direction [{\it synchronism}]. In such a case the elementary magnets would be oriented in the same direction, i.e. they would prefer to be parallel. He gave a result for the energy gain $\Delta E$ when being parallel

\begin{equation}
\Delta E= \frac{e^2}{a}\Big(1 - \frac{1}{\sqrt{1+k(\frac{r}{a})^2}}\Big)  \label{shcottkyresult}
\end{equation}
where $e$ is the electron charge, $a$ the distance of the atoms nuclei in the solid and $r$ the radius of the electron around the nucleus. The value of $k$ depends on the geometry of the electron orbits. If they  are above each other $k=4$ if they are in the same plane $k$ is about 2. No explicit calculation is given. A more detailed discussion has been published recently\cite{foho}. 

When one now considers two neighboring atoms in the solid, they may be arranged besides each other in the same plane or above each other. The clockwise or counter clockwise rotation of the electron corresponds to a parallel or antiparallel orientation of the magnetic moment created by the moving charge. During their rotation the two electrons approach each other and the minimal distance defines the energy necessary for the continuos rotation.
This minimal distance depends on the phase difference of the circulation. In each case the electrons are synchronized [that means the phase adjusts itself] in such a way that the minimal distance is as large as possible.
It turns out that in each case either above or beside each other the energy necessary is smallest if the rotation is in the same direction and the magnetic moments are parallel.

In consequence their induced elementary magnetization is parallel. The energy difference $\Delta E$ to the antiparallel configuration turned out to be of the order of the Curie temperature if one inserts reasonable values for the respective parameters of the solid ferromagnets ($T_c\approx \Delta E/k_B$). He called this interaction {\it roto-active directivity} [{\it rotoaktive Richtwirkung}]. Finally the reason for this behavior of the electrons remained unknown.


However Schottky \cite{schottky1922} suggested that this interaction is of electrostatic nature and based his considerations on an already general principle the {\it synchronismus} used  in the planetary motion of electrons in atoms or between atoms \cite{lande1921}. The idea was that in crystals the structure results from the electrostatic attraction and repulsion between the atomic nucleus and the (planetary like) moving electron. Land\'e wrote:
``The regular structure of the monatomic crystals also leads to the conclusion that all particles are ordered in their electron movement phases.". This phenomenon was called {\it lattice synchronism} ({\it Gittersynchronismus})  and Schottky used it now for the ordering of the magnetons.  


\begin{figure}[h]\centering \includegraphics[height=0.4\textwidth]{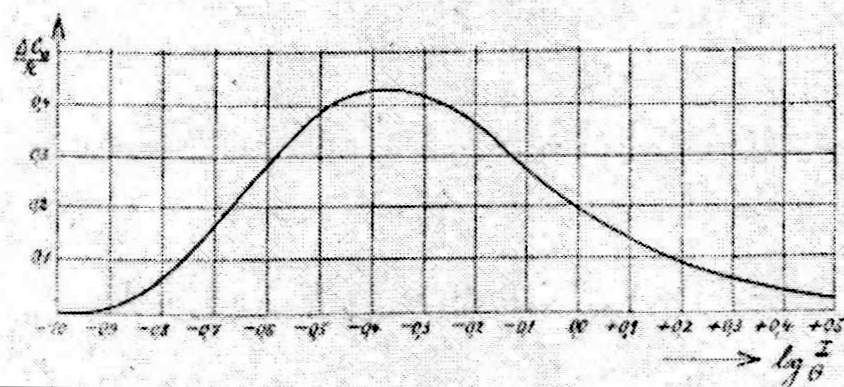}
\caption{\label{anomaly}The specific heat of a system with two different energy levels. The peak is called Schottky anomaly. Fig. from Ref. [\refcite{schottky1922}].}
\end{figure} 

As already mentioned, Schottky's paper is  nowadays know for the calculation of the specific heat in a two-state system. The presence of this two states differing in energy lead to a peak in the specific heat at low temperature, named {\it Schottky anomaly} (see Fig. \ref{anomaly}) which is given by 
\[C/k_B=\frac{(\Theta/T)^2e^{-\Theta/T}}{(1+e^{-\Theta/T})^2}=\Big[\frac{(\Theta/T)}{\cosh(\Theta/2T)}\Big]^2\] 
As late as 1937 the specific heat of the Ising model was calculated by Bitter\cite{bitter37} without recognizing that it shows an anomaly as calculated by Schottky. In fact the Ising chain in zero external field is dual to the two state model (see Sect. \ref{dual} below).  

\section{A hard task for Ising \label{hard}}

{\it The more complicated the system considered, the more simplified must its theoretical description be. \dots
A good theory of complicated systems should represent only a good "caricature" of these systems, exaggerating the properties that are most difficult, and purposely ignoring all the remaining inessential properties.}[Frenkel 1946 cited in 1962 by Tamm\cite{tamm}]

After the results of Schottky's ideas on a possible interaction were published  Lenz  proposed Ernst Ising to find out if his idea from 1920 can be formulated in a model showing a ferromagnetic phase for his  dissertation. One may cite here Heisenberg who said\cite{heisenberg58}: ``What we observe is not nature itself, but nature exposed to our method of questioning." How do we find out if a magnetized phase exists? At the end of the year 1922 Ernst Ising started with his thesis to answer this question.

In principle it was clear how to attack the problem. In fact it looked easy since the formalism was already developed, namely one had to calculate the partition function of the system in an external magnetic field. Then one could look if a magnetized phase could be found by deriving the free energy  by the magnetic field and setting the external magnetic field to zero. 

Indeed Ising followed\cite{thesis} the standard path known in statistical physics developed by Boltzmann and/or Gibbs. The magnetization $\frak J$ (his notation is used) is obtained by deriving the free energy (a function of temperature $T$ and external field $\frak H$) after the external field. The free energy in turn is calculated from the partition function $Z$ thus 
\begin{equation}
\frak J=\frak m \frac{\partial}{\partial\alpha}\log Z(T,\alpha) 
\end{equation}
with $\alpha= {(\frak m \frak H)}/{k_BT}$, $\frak m$ the elementary magnets, which can only be parallel or antiparallel and $k_B$ the Boltzmann constant.

\begin{figure}\centering
 \includegraphics[height=0.12\textwidth]{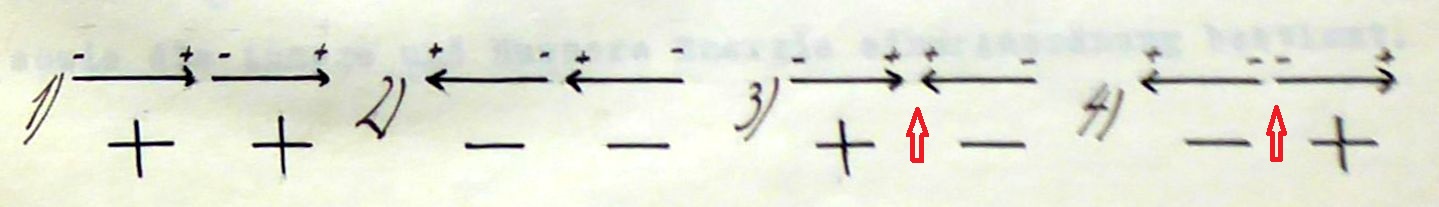}
\caption{\label{konf} Ising's representation of the microstates: Antiparallel configurations cost energy (red arrows), the direction of the arrows are irrelevant and therefore Ising replaced them by + and -.  . The picture is taken from the original Ising's thesis, red arrows are added by the author.}
\end{figure}

The partition function contains the number  $\cal N$ of microstates multiplied by their Boltzmann's weight given by the energy of the microstate. It has to be noted that no Hamiltonian could be written down for the microstates due to the vague microscopic model of Schottky. 
In order to define the microstates of the sytstem Ising introduced the notation shown in Fig. \ref{konf}.
He counted the elementary magnets in (against) the external field direction by $\nu_1$ ($\nu_2$) and assigned pluses and minuses to them. The places where two elementary magnets are antiparallel he called `energy places' (red arrows) since this configuration with respect to the parallel one needs the additional energy $\epsilon$. There are $\sigma$ energy places counted. These three numbers $\nu_1$, $\nu_2$ and $\sigma$ define now the microstates and one has to find their number  
\[{\cal N}(\nu_1,\nu_2,\sigma) .\]
Ising starts from  $\nu_r(m)$ the number of ways in which the integer $m$ may be represented as a sum of $r$ integer numbers.  The result is the 
binomial coefficient
\[\nu_r(m)= \binom{m-1}{r-1} . \] 
Now he applied this to positive and negative elements in the chain and observed the configuration of the last boundary element [$\delta=0,1$] if it is parallel or antiparallel to the first boundary element. This gives the result for the number of microstates
\[{\cal N}(\nu_1,\nu_2,\sigma=2s+\delta)= \binom{\nu_1-1}{s}\binom{\nu_2-1}{s+\delta-1} +\binom{\nu_2-1}{s}\binom{\nu_1-1}{s+\delta-1} .\]
The energy of the system is given by the difference in number of elementary magnets in and against the external field $(\nu_1-\nu_2)$. The number of energy states is $\sigma=2s+\delta$, where $s$ is the number of the `internal' energy states.
Then the partition function $Z(n)$ for $n$ elementary magnets in the chain [$\alpha\sim \mathfrak H$; $\beta\sim\epsilon$] reads
\[Z({n})=\sum_{\nu_1+\nu_2=0}^{n}\sum_{s=0}^{\infty}\sum_{\delta=0}^1 {\cal N}(\nu_1,\nu_2,\sigma=2s+\delta)e^{(\nu_1-\nu_2)\alpha-(2s+\delta)\beta} .\]
Ising now proceeded with a clever trick. First he introduced an auxiliary variable $x$ and defined a function $F({x})$
\[F({x})=\sum_{{n}=0}^\infty Z({n}){x}^{n}\]
by adding up the partition functions for all different length $n$ of chains and calculated the sum over $n$:
\begin{align*}
{F}(x)=&\sum_{s=0}^\infty\sum_{\delta=0}^1 B^{2s+\delta}
\left[
\left( \sum_{\nu_1=0}^\infty\binom{\nu_1-1}{s}(xA_1)^{\nu_1}\right)
\left( \sum_{\nu_2=0}^\infty\binom{\nu_2-1}{s+\delta-1}(xA_2)^{\nu_2}\right)
\right.\\
&\qquad\quad\left.+
\left( \sum_{\nu_1=0}^\infty\binom{\nu_1-1}{s+\delta-1}(xA_1)^{\nu_1}\right)
\left( \sum_{\nu_2=0}^\infty\binom{\nu_2-1}{s}(xA_2)^{\nu_2}\right)
\right] .
\end{align*}
The number of elementary magnets $n$ has disappeared and was replaced by the auxiliary variable $x$. In order to get back the partition function for length $n$ he now expanded the function in powers of $x$ 
and looked for the coefficient of $x^{n}$ for large ${n}$. This gives back the partition function for the length $n$:
\[{F}({x})=\frac{2{x}[\cosh\alpha-(1-e^{-\epsilon/k_BT}){x}]}{1-2\cosh\alpha\cdot {x}+(1-e^{-2\epsilon/k_BT}){x}^2} = \sum_{\mathbf{ n}=0}^\infty\left(\frac{a_1}{w_1}{w_1}^{\mathbf{n}}+ \frac{a_2}{w_2}w_2^{\mathbf{n}}\right){x}^\mathbf{n}\] \\
with $w_1=\cosh \alpha+\sqrt{\sinh^2\alpha +e^{-2\beta}}$ and $w_2<1$. In this way he recovered the partition function for the chain in the {\it thermodynamic limit} by neglecting the terms $w_2$ which decays for large $n$.

The magnetization $\frak J$ is then given by the derivative of the partition function $Z$ with respect to the field $\alpha= {(\frak m \frak H)}/{k_BT}$ 
\[
\frak J=\frak m \frac{\partial}{\partial\alpha}\log Z=\frak m n\frac{\partial}{\partial\alpha}\log {w_1}\]
with the result
\[{\frak J=\frak m n\frac{\sinh\alpha}{\sqrt{\sinh^2\alpha+e^{-2\beta}}}} .\]
From this one concludes that in the case of zero external field, $\alpha=0$, there is no finite magnetization at temperatures larger than zero.
In consequence there is {\bf no ferromagnetic phase at any finite temperature} in the linear chain.

Ising tried of course to apply his method to more complicated cases in order to see if he could calculate the partition function in higher dimensions in presence of an external magnetic field. This is even today not possible, thus he could only extend his calculations by making further simplifications and/or approximations in order to proceed to results.
So he looked for the cases of
\begin{itemize}
\item the ladder or double chain,
\item the chain with next nearest neighbor interaction, and
\item a chain with more than two possible directions of the elementary magnets at one place.
\end{itemize}
None of these extensions made a ferromagnetic phase possible and so Ising ended his thesis with the {\it supposition} that also in the three dimensional case the model does not lead to a ferromagnetic phase at low temperatures.

\section{The evaluation of Ising's result}

\subsection{Lenz approval of Ising's thesis}

Of course Lenz was also disappointed by this result and as can be seen by his approval of the thesis\cite{lenzappr}:
``The more than hundred-year-old theory of ferromagnetism has come to a certain formal conclusion in the papers of P. Weiss. These makes it possible to express the thermal behavior of a ferromagnet in a formal way. However, the Weissian developments can not be regarded as a satisfactory theory, since their foundation, the so-called molecular field, themselves has the character of an auxiliary approach. A satisfactory theory must be based on the behavior of the atoms of a solid body, for which  Bohr's theory provides the necessary clues. Following a theory of thermal behavior of the \underline {paramagnetic} salts that satisfies these requirements, I have suggested to the candidate to extended these ideas to ferromagnetism. The necessary rather complicated probabilistic considerations have been carried out by the author with remarkable skill. They led to the conclusion that ferromagnetism is not realized on the path taken, and the reasons for this are being discussed. Since a change of the basic ideas about the atomic properties and their interaction is not an option, the question arises as to whether the ferromagnetic state can at all be regarded as a state of thermal equilibrium. The study of this possibility, however, would have far exceeded the scope of this thesis.

The overall negative result must be viewed as an important statement for the continuation of the theory. The work (\sout{deserves} \sout{the}) also provides some new methodological aspects and deserves the predicate: ,,very,,good"  Lenz"

Cancellation by Lenz,  he also first\cite{reich} gave the mark ,,good'' but changed it into ,,very good'' as seen by the quotation marks. He supported his decision  by  the remark on the new methodical aspect of Ising's work. Indeed one finds later citations of Ising's paper pointing to his combinatorial methods. One was published in 1939 by A. H. Mood in the paper\cite{mood} based on his thesis and the other one in 1942 by  T. S. Chang and C. C. Ho \cite{chang1942}. T.S. Chang was a student of Ralph H. Fowler working in statistical physics at the University of Cambridge, United Kingdom. Fowler and Guggenheim published in 1939 the well known textbook {\it Statistical Thermodynamics} where the combinatorial methods are used explaining order-disorder phase transitions in alloys.
 
Surprisingly Lenz did not explicit notice that Ising has shown that the Curie-Weiss theory is wrong in the one-dimensional case. This failure of the mean field approximation was corrected ten years later by Bethe (see Sect. \ref{impact}).


 \subsection{The reduction made in the publication}
 
 After finishing his thesis Ising submitted on December 9, 1924 a short summary of his thesis to Zeitschrift f\"ur Physik\cite{Ising25}. In order to be short he drastically reduced the references.  Thus one may count about 18 papers in the thesis and only 3 papers in the publication. Most important he did not cite the oldest  and the newest reference namely those which concerns Kirwan's\cite{kirwan} and Schottky's paper\cite{schottky1922}. He instead just mentioned:
 ``We assume that the internal energy is smallest when all elements are in the same direction.'' Both references in fact dropped out of the history of ferromagnetism. Thus only two references of Weiss\cite{weiss2,weiss3} to the molecular field theory (in French and German) and Lenz's paper\cite{lenz1920} remained.
 
Although the result of Ising's publication is generally seen as negative - no magnetic phase in the three dimensional system - on the other hand it is positive in the formulation of a model and by calculating an example of the failure of the mean field theory in the case of one dimensional systems.  Fortunately the negative result of the model was published leading to an unforeseen success in the following decades. The tendency that negative results nowadays are disappearing in most disciplines has been studied by bibliometric methods\cite{fanelli}.
 
 \subsection{Naming the model}

There is on and off a discussion how to name the model which was suggested by Wilhelm Lenz and mathematically defined and solved in one dimensions by Ernst Ising.  In the first review on this model  Brush \cite{brush} noted: ``Lenz himself apparently never made any attempt later on to claim credit for suggesting the model, and even his colleagues at Hamburg University were not aware of his contribution [Professor Dr. H. Raether in a private communication to Brush]." However Brush decided to name the model ``Lenz-Ising Model'' possibly due to the correspondence with Ising. But in 1971 in Vol 2 of {\it The Kind of Motion We Call Heat} (p.12) he wrote\cite{brush2}:\vspace{0.02cm}

{\it ...it is usually futile for historians of science to try to change established nomenclature,...}\vspace{0.02cm}

He added in footnote 21:\vspace{0.02cm}

{\it My attempt to add the name of Wilhelm Lenz to the ``Ising Model'' in statistical mechanics [Rev. Mod. Phys. {\bf 39}, 883 (1967)] has been a failure.}\vspace{0.02cm}

Indeed looking at citations in the publications dealing with this model almost all authors name it ``Ising model''. A rare exceptional case is the publication\cite{kramers1939} by H. A. Kramers at the magnetism conference in Strasbourg 1939 but  later in  1941 he cited the model as Ising model too.
However others kept the naming  in later publications [referring to Brush's review, but without knowing his  withdrawal]. In the relevant scientific papers looking for its solution the model is always cited as {\bf Ising Model}.

 \section{Ising leaving Hamburg and his new phase of life until 1930}

 \parbox{0.3\textwidth}{
\includegraphics[height=5.3cm]{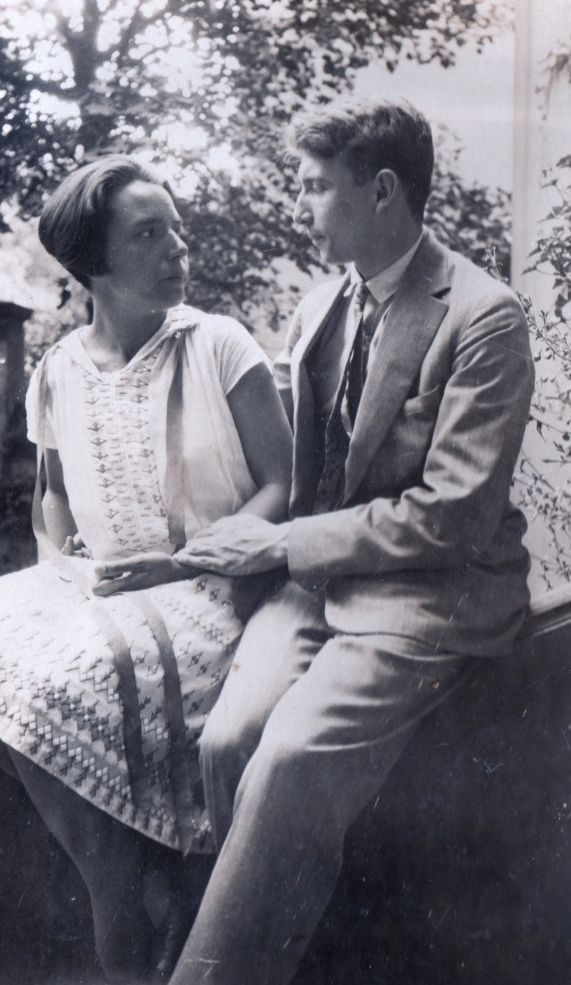}\\ Johanna and Ernst {\tiny\copyright Tom Ising}}
\hspace{0.03\textwidth}
\parbox{0.6\textwidth}{\vspace{-0.5cm} \begin{enumerate}
 \item[1925]  Job at the patent office of the Allgemeine Elektrizit\"asgesellschaft (AEG)
  \item[1926] Ernst Ising meets Johanna Ehmer
  \item[1927] Teacher at the boarding school in Salem
 \item[1928] Studying  philosophy and pedagogy at Berlin University 
 \item[1930] State examen; teaching position Strausberg near Berlin as {\it Studienassessor} [higher civil service]; marrying Johanna Ehmer who held a doctorate in economics
 \item[1932] Ernst transfered from Strausberg to Crossen/Oder 
 \end{enumerate}} \vspace{0.5cm}

After Ernst Ising had finished his thesis and his publication he left Hamburg in 1925 and went to Berlin. He  found a job in the patent office of the General Electric Company [Allgemeine Elektrizit\"asgesellschaft (AEG)]. There in Berlin on June 6, 1926 he met Johanna Ehmer at a Sunday outing of the Pacifistic Student organization\cite{johannaWalk}.

Johanna was born in 1902 in Berlin, her father being a drugstore owner her mother a bookkeeper with various businesses. After graduation from the gymnasium Johanna enrolled in mathematics and physics at the University of Berlin since in these subjects she was excellent in school. However after three semesters she switched to economics and proceeded  her studies at the College of Commerce (Handelshochschule) where she earned the commercial teacher's diploma in spring 1924. In 1926 she finished her thesis supervised by Charlotte Leubuscher, who was the first woman in Germany habilitated in humanities.
Looking for a position, Johanna  became Research Assistant at the Department of Economics assigned to Professor Franz Eulenburg. 

 Ernst Ising was not satisfied by his job and decided to switch to teaching. Therefore he resigned and found a job at the boarding school in Salem (South Baden) in 1927 and took some classes at the University of Berlin in order to finish the the Prussian State examination in 1928. After two years more of teacher training he became ``Studienassesor'' (the civil servant position at a high school) and on December 23, 1930 (six years after finishing his thesis) Ernst and Johanna married. Ernst had to teach in Strausberg near Berlin and later in fall 1932 in Crossen/Oder far east of Berlin.




\section{The model after the thesis until its modern formulation in 1930}

\subsection{Herzfeld 1925}

Karl Herzfeld was the first at the Annual Meeting of  German Physicists in Danzig to refer  \cite{herzfeld} to Ising's publication in the year of its publication. Herzfeld had studied in Vienna and his 
thesis was supervised by Hasen\"ohrl. After five years in Munich he became in 1925 an extraordinary professor there. A year later he went as a visiting professor to the United States and remained there until his death in 1978 \cite{herzfeldbio}. Herzfeld gave a review with the title {\it Molecular- 
and Atomic Theory of Magnetism} in which he also refered to Schottky's idea of synchronism and roto-active interaction citing several publications of  Land\'e \cite{lande1921}, Born, Heisenberg\cite{bornheisenberg} und Nordheim \cite{nordheim}. Herzfeld was not convinced in the argumentation concerning the interaction assumed. He criticized the restriction to only nearest neighbor interaction and argued that in such a case the energy is the same  irrespective of only one dipole or a whole group of dipoles lies antiparallel.

\subsection{Pauli and the New QM}

The ``old" Bohr-Sommerfeld quantum mechanics used three quantum  integer numbers for the classical orbital motion of an electron corresponding to characterize the planetary-like path of the electron around the nucleus. 
This model could explain the simplest situation in an atom with one electron but failed in more complicated cases e.g when more than one electron is present or when external fields are present. 

Pauli was struggling to understand atomic magnetism.  He tried to explain the anomalous Zeeman effect but did not succeed \cite{pauli0}. He used the neighborhood to Copenhagen and worked on sabbatical at Bohr's institute from September 1922 to  October 1923. On  17 January 1924 he submitted his application for habilitation 
and on  February 23, 1924 he already held his inaugural lecture \cite{reich}. 
In the rest of the year 1924 he developed further his idea of a fourth quantum number for the electron\cite{pauli1} - his ``classically not describable two-valuedness in the quantum mechanical  properties of the valence electron" (for the non-classical character see W. Pauli's Scientific Work by Ch. Enz\cite{enzPauli} p. 29).  It was in the same issue of {\it Zeitschrift f\"ur Physik} received seven days earlier as, and printed 120 pages behind of Ising's paper.  Shortly later on January 16, 1925  the exclusion principle \cite{pauli2} was received at the journal.  
This principle says: ``There can never be two or more equivalent electrons in the atom for which the values of all quantum numbers ... coincide in strong fields. If there is an electron in the atom for which these quantum numbers (in the outer field) have certain values, this state is {\it occupied}.... We cannot give a more detailed explanation for this rule, but it seems to be very natural on its own." No relation between Ising's model for explaining macroscopic ferromagnetism and Pauli's new quantum degree of freedom of the electron to explain the microscopic magnetism in the atom was seen at that time.

Pauli describes his motivation in the lecture\cite{lecturePauli} given in Stockholm after the award of the Nobel Prize in Physics in 1945: (1) ``One was an effort to bring abstract order to the new ideas by looking for a key to translate classical mechanics and electrodynamics into quantum language which would form a logical generalization of these.''
(2) The second was ``a direct interpretation, as independent of models as possible, of the laws of spectra in terms of integral numbers, following, as Kepler once did in his investigation of the planetary system, an inner feeling for harmony.''

Already at that time Pauli was in close contact with Werner Heisenberg. In the years 1925 to 1928 forty letters from Heisenberg to Pauli are archived, but unfortunately Pauli's answers are not conserved. C. F. von Weizs\"acker remembers\cite{PBW1} a remark of Heisenberg: ``I never have published a work without Pauli reading it first. If he said it was wrong, it could still be right; but then caution was needed." Both Pauli and Heisenberg shared the doubts on the classical picture of the atom as planetary system. Thus Heisenberg  wrote to Pauli on June 9, 1925: ``It is really my conviction that an interpretation of the Rydberg formulas in lines of circles and elliptical orbits in classical geometry does not have the slightest physical sense and all  my poor calculations go to the point to kill completely the term orbits that cannot be observed anyway, and to replace it properly." 
Twenty days later Heisenberg submitted his famous paper\cite{heisenberg1} on the new interpretation of quantum mechanics for publication. 

The same reservation had Pauli against a classical interpretations of the new degree of freedom of the electron. A first interpretation given by Ralph Kronig  was not submitted due to the criticism of Pauli and a contribution\cite{kronig} to the spin appeared later. It was Goudsmit's and Uhlenbeck's paper\cite{goudsmit} of the year 1925 on the spinning electron which explained the fourth degree of freedom. With the rotating charged electron a magnetic moment was connected. The problem with the size of this moment (the electron had to rotate with a velocity at the circumference  higher than the velocity of light) was solved later by taking into account relativistic effects\cite{thomas} by Thomas.

Meanwhile also the new quantum mechanics was further developed by Born, Jordan and Heisenberg and Pauli contributed\cite{pauli26} to that development by solving the hydrogen atom within the new formalism. Moreover Schr\"odinger formulated another interpretation\cite{schroedinger} of quantum mechanics - the wave mechanics in contrast to Heisenberg's matrix mechanics and showed its equivalence.``A new mathematically essential more convenient access to the field of quantum mechanics''  as Heisenberg\cite{heisenberg1926} agreed.

 \subsection{The New QM and Heisenbergs exchange interaction}
 
As early as on November 4, 1926 Heisenberg communicated some ideas about ferromagnetism to Pauli. He wrote (underlining due to the original): ``The idea is this: In oder to use Langevin's theory of ferromagn., one has to assume a strong coupling force between the spinning electrons (\underline{only} \underline{these} turn around). As with helium, this force should be supplied indirectly by the resonance. I think one can generally prove: parallel positioning of the spin vectors always gives \underline{smallest} energy. The energy differences that come into consideration are of the \underline{electrical} order of magnitude, but decrease \underline{very} rapidly with increasing distances. I have the feeling (without even knowing the material remotely) that this could in principle be sufficient for an interpretation of ferromagnetism." 

But it needed one and a half year more until he was able to bring these ideas into the mathematical formalism of the new quantum mechanics. An important step was to include the symmetry property of an system of electrons. This was done almost simultaneously by Heisenberg\cite{heisenberg1926} (June 11) and 
Dirac\cite{dirac1926} (August 26) who explained: "If the positions of two of the electrons are interchanged, the new state of the atom is physically indistinguishable from the original one.'' A second  important step further was the introduction of the spin by 2$\times$2  matrices into the formalism of quantum mechanics by Pauli\cite{pauli1927}.

 On May 20, 1928 the paper of Heisenberg was received by the journal Zeitschrift f\"ur Physik. In his one sentence abstract he proclaimed\cite{heisenberg28,heisenberg28e}: ``{\sc Weiss}'{\sc s} molecular forces will be attributed to a quantum-mechanical exchange phenomenon, and indeed, it will be treated as the exchange process that was successfully enlisted in recent times by {\sc Heitler} and {\sc London} in order to interpret homopolar valence forces.''
 He based his theory  on the {\sc Pauli-Fermi-Dirac} statistic property of the electron and the {\sc Pauli} principle. The eigenfunctions of the quantum mechanical system are constructed by allowing electrons to be exchanged between neighboring atoms. Then the energy depends on the ``very important
constant that enter into the perturbation calculations is the purely {\sc static} [Coulomb] interaction'': 
\begin{equation}
J_E=\int d\tau_1d\tau_2\ldots d\tau_{2n}(\psi^1_1)^2(\psi^2_2)^2\ldots(\psi^{2n}_{2n})^2\big(\sum\frac{e^2}{r_{kl}}+\sum\frac{e^2}{r_{\kappa\lambda}}-\sum\frac{e^2}{r_{k\lambda}}\big)
\end{equation}
where $k$ and $l$ denote the electrons and $\kappa$ and $\lambda$ the atoms and $\psi^\kappa_k$ the corresponding spatial part of the wave function. This energy is the Coloumb energy and Heisenberg argues: ``due to their smallness, we can leave  the magnetic interaction outside of consideration.'' In fact this establishes within the formalism of the new quantum mechanics, what has been suggested by Schottky within the old quantum mechanics\cite{foho} (see section \ref{schottky}). Due to the Pauli principle the wave functions lead to a charge distribution of the electrons which minimizes the Coulomb energy for two neighbouring electrons with parallel spin. Thus the vague synchronism was replaced by new nonclassical physical principles.  

With this Heisenberg calculated the partition function in Gaussian approximation  in order to find the Curie temperature $\Theta$. A finite value results in this approximation
\begin{equation}
\Theta= \frac{2J_0}{k_B\big( 1-\sqrt{1-8/z}\big)}
\end{equation}
where $k_B$ is the Boltzmann constant and $z$ the number of nearest neighbours. From this expression for $\Theta$ Heisenberg concluded that in three dimensions a phase transition is possible only for $z>8$. This was considered to confirm the result of Ising since no real finite transition temperature exists for the chain where $z=2$. The value for the magnetization turned out to be essentially in agreement with Weiss's theory.

A final overall  picture on the microscopic level for the description of the electron was given in 1928 by Dirac\cite{dirac1928}. He wrote: I ``would like to find some incompleteness in the previous methods of applying quantum mechanics to the point-charge electron such that, when removed, the whole of the duplexity phenomena [the spin of the electron] follow without arbitrary assumptions.'' In an  interview around 1982 by Hund\cite{hundDirac} he further explained:   
  ``that forced me into a different kind of equation [now the famous Dirac equation] and this different equation brings in the spin of the electron. It was very unexpected to me to see the spin appearing in that way. [spin means both moment of momentum and magnetic moment Hund adds] I thought one would have a satisfactory theory without spin and then one would proceed to a more complicated theory.'' Thus the spin and magnetism of the electron turned out to be an inevitable property emerging out of quantum theory and special relativity\cite{berry}.
 
Now all the necessary microscopic details have been developed including an appropriate mathematical method to describe the non-classical two-valuedness and more than five years after Ising had formulated his model it became possible to set up a mathematical description by defining a Hamiltonian for it. 

\subsection{Pauli and Heisenberg on the Ising model and the thermodynamic limit}

\subsubsection{Heisenberg's comments on Ising's model 1928 \label{heiscom}}
Ising tried to set up a model in more than one dimension, but this was based in fact on simplifications and/or approximations in the calculation of the partition function, which did not lead to ferromagnetism and in fact not to a true three-dimensional model. After the new developments of the microscopic basis of possible macroscopic magnetism  Heisenberg \cite{letterH1928}  criticized Ising's conclusion:

``The whole question [about ferromagnetism] seems important to me because of the similarity between my model and that of Ising. According to my current view, Ising would have gotten ferromagnetism, if he had accepted enough neighbours (about $z \le 8$) [Authors remark: Even if one accumulates sheets of planar Ising planes to a cubic model it would have only $z=6$]. After the argument that Ising has published for the spatial model, it seems to me at all, as if he had not understood his work. As long as one does not go over to the limit $n \to \infty$, it is trivial that for $H = 0$ also $J = 0$, and in fact with parallel position of all magnets [footnote: {One just applies Langvin's theory for the magnetic rod as a whole!}] becomes $J$ prop. {$\mathfrak{Tg}$}(n$\alpha$). In Ising's linear chain parallel arrangement of all spin vectors occurs if $e^{-2\epsilon/kT} \sim \frac{1}{n}$ holds, then $T=\Theta\sim \frac{2\epsilon}{k\lg n}$; the Curie point $\Theta$ is thus dependent on $n$ here and goes to zero as $n$ increases. For the wild spatial model of Ising, however, would be $\Theta\sim \frac{2\epsilon n^{3/2}}{k\ln n}$, $\Theta$ would go to infinity with increasing n, i.e. parallel arrangement of all magnets would always exist. The fact that Ising conceived this model as an argument against ferromagnetism seems to me a sign that he has not understood his own work remotely. What do you think about it?"

Unfortunately  Pauli's answers are lost. Nowadays we know that the spatial dimension is a relevant parameter for the existence and nonexistence of a phase transition and not the numbers of nearest neighbours. Heisenberg argues with his result obtained in Gaussian approximation. Ising's result for the chain is exact and holds in the thermodynamic limit this includes extensive quantities which are proportional to $n$. 

In a finite chain the probability that all spins are parallel is indeed finite and given by
\begin{equation}
P=\frac{1}{2}\Big(\frac{e^{\epsilon/k_BT}}{\cosh (\epsilon/k_BT)}\Big)^{n-1}
\end{equation}
which is below some temperature almost  $1/2$ and leads to a `transition temperature' which goes to zero with $n\to \infty$. The ordering in finite chains has been studied\cite{haenggi}  in 2005. There the authors  ``have derived a criterion that a finite Ising chain exhibits the ferromagnetic like behavior. According to it, the transition time between the fully magnetized chain states must exceed the measuring time, and the average number of domain walls must be much less than 1. These conditions hold, i.e., a finite Ising chain does display a ferromagnetic like order on the measuring time scale, if the temperature is sufficiently small.''


\subsubsection{Pauli at the Solvay conference in 1930}

 \begin{figure}[h]
\centering\includegraphics[width=\textwidth]{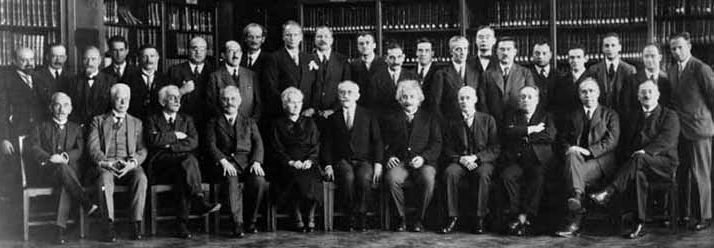}
\caption{\label{solvay30} Solvay conference 1930. Standing: 
E. Herzen, E. Henriot, J. Verschaffelt, C. Manneback, A. Cotton, J. Errera, { O. Stern}, A. Piccard, W. Gerlach, C. Darwin, { P.A.M. Dirac}, E. Bauer, P. Kapitsa, L. Brillouin, H. {\bf A. Kramers}, P. Debye, {\bf W. Pauli}, J. Dorfman, { J. H. Van Vleck}, E. Fermi, { W. Heisenberg}; seated in front:
Th. De Donder, P. Zeeman, { P. Weiss}, A. Sommerfeld, M. Curie, P. Langevin, A. Einstein, O. Richardson, B. Cabrera, N. Bohr, W. J. De Haas. Pauli presented the Ising model in his talk and eleven years later Kramers with Wannier proofed Pauli's conjecture and calculated the transition temperature for the two-dimensional Ising model. From Ref. [\refcite{marage}].}
\end{figure}  

 Pauli was invited to present a review talk \cite{pauli1930} at the 6th Solvay conference in Brussels from  October 20 to 25 with the title {\it Les Th\'eories Quantiques du Magn\'etisme: L'\'electron Magn\'etique} [{\it The Quantum Theories of Magnetism: The Magnetic Electron}].
 
He expounded  how the new development on the microscopic level in quantum mechanics as the spin, the new formulation of by Heisenberg, Schr\"odinger and Dirac and the exchange interaction contributed to the understanding of magnetism and especially of ferromagnetism. With this he was able to formulate the Hamiltonian function $\mathcal H$ for the Ising model (he considers the chain)
\begin{equation} {\mathcal H}=-A\sum_k(\sigma_k,\sigma_{k+1}) \label{Hising} \end{equation}
where the $\sigma_k$ is the spin matrix corresponding to the direction of an possible external field and might be chosen to be in the $z$-direction. The strength of the exchange interaction of  two neighboring electrons indexed by $k$ and $k+1$ is given by $A$. The sum is understood over all neighbours. This has to be distinguished from the Heisenberg model
\begin{equation} {\mathcal H}=-A\sum_k({\vec{\sigma}}_k,{\vec{\sigma}}_{k+1})  \label{Hheisenberg} \end{equation}
 where the components of the vector contain all three spin matrices.  
   
Pauli expressed his expectation that Ising's model could lead to a phase transition in higher dimensions.
He stated in the discussion\cite{pauli1930}  of the Hamiltonian \eqref{Hheisenberg}: ``There is in fact a very close relationship between the problem of Ising and the one we have just treated" , and explains: ``In Ising’s calculation developed from the point of view of the old quantum mechanics, the components of $\sigma_i$ that are perpendicular to the field are considered to be zero, whereas in the new quantum theory these components do not commute with the components in the direction of the field." But he immediately suspects for the Ising model \eqref{Hising}: ``Irrespective of this difference, it is quite likely that an extension of the theory of Ising to the case of a lattice of three dimensions would yield ferromagnetism even from the classical point of view".


\section{Ising as a teacher from 1933 until the November pogrom in 1938}

 \parbox{0.55\textwidth}{
\includegraphics[width=0.55\textwidth]{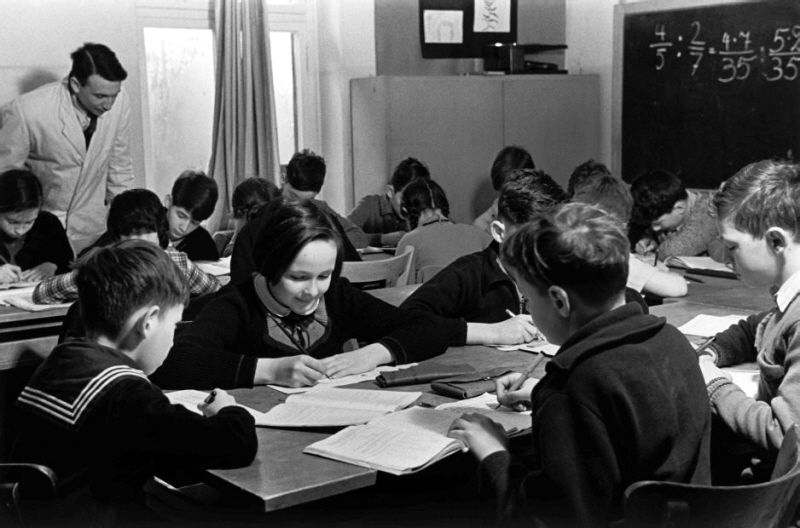}\\ Ernst teaching in Caputh\\
{\tiny\copyright Herbert Sonnenfeld, J\"udisches Museum Berlin}}
\hspace{0.04\textwidth}
\parbox{0.39\textwidth}{
 \begin{enumerate}
 \item[1933]  March 31 Dismission
 \item[1934] Teacher in Caputh
 \item[1937] Headmaster of the school
 \item[1938] November 10 Destruction of the school during the Pogrom
   \end{enumerate}}\vspace{0.5cm}\\

The {\it Paradise Lasted A Year And A Half}\cite{johannaWalk} for  Ernst and Johanna until Hitler came to power in March 1933 by the Enabling Act. The Enabling Act allowed Hitler and his Cabinet to rule by emergency decree for four years. New antisemitic laws followed which forced all Jewish scientists (but not only them) to resign their position and/or leave Germany.
On April 7, 1933, the {\it Law for the Restoration of the Professional Civil Service}  (Gesetz zur Wiederherstellung des Berufsbeamtentums) came into force\cite{wolff1993}, which concerned Ernst since he was dismissed  from his teaching post in public schools. On May 1, 1933, the University of Hamburg took the {Vow of allegiance of the Professors of the German Universities and High-Schools to Adolf Hitler and the National Socialistic State}. 

From one day to the next, all government jobs held by Jews were clearly at risk. At the universities, all of them state institutions, waves of dismissals began promptly\cite{pt2014} with all its consequences on the existing network of physicists. These consequences for German and Austrian science and its future were estimated\cite{waldinger} and it was indicated  that  the reduction in output in departments with dismissals persisted at least until 1980.

\begin{figure}[h]\centering
\includegraphics[width=0.41\textwidth]{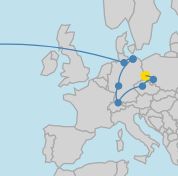}\hspace{0.05\textwidth}\includegraphics[width=0.38\textwidth]{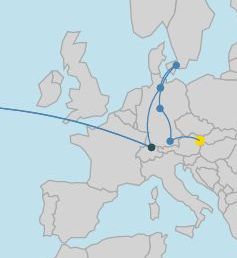}
\caption{\label{moves}Career path\cite{careerPT} of Nobel physicists starting from the  birth place (yellow) (a) for Stern until the  escape 1933 and  (b) for Pauli until the move 1940 to the United States  \copyright Physics Today}
\end{figure}

Otto Stern resigning from his post at the University of Hamburg, found refuge in the city of Pittsburgh becoming a professor of physics at the Carnegie Institute of Technology.  Wolfgang Pauli had already left the university in 1928, when he was appointed Professor of Theoretical Physics at ETH Zurich in Switzerland. The German annexation of Austria in 1938 made him a German citizen, which became a problem for him in 1939 after the outbreak of World War II. In 1940 he tried in vain to obtain Swiss citizenship, which would have allowed him to remain at the ETH. Pauli instead moved to the United States in 1940, where he was employed as a professor of theoretical physics at the Institute for Advanced Study. Not until 1946, after the war, he became a naturalized citizen of the United States and subsequently returned to Zurich.

Many physicist, who worked in the field of magnetism and  had published on the Ising model, had to emigrate, especially the young generation of post docs and assistants. When J. J. Thomson, who discovered the electron, opened a meeting of physicists in Oxford in 1936, he opened it with the words\cite{nachmansohn} ``Heil Hitler! Thanks to Hitler we have today with us Franz Simon, Kurt Mendelssohn [great authorities in the field of low-temperature physics], Rudolf Peierls, and many others''.
   
Leubuscher, who supervised Johanna's thesis, half-Jewish lost in 1933 the habilitation and emigrated to Great Brittain where she  was in the network\cite{oertzen} of the  British Federation of University Women (like Lise Meitner, Marie Jahoda, Charlotte B\"uhler). Johanna could stay at the  College of Commerce [Handelshochschule], not a state institution in Berlin, where she privately worked for Prof. Franz Eulenburg. He  was dismissed as a Jew not until 1935 because of the large amount of doctoral candidates he had to care for. He died in 1943 under unexplained circumstances  in the prison of the Gestapo.

Also Jewish pupils were dismissed from public schools and gathered in special Jewish schools. One of such institution was the Landschulheim (boarding school) Caputh, founded already in 1931 not only for Jewish children by Gertrud Feiertag\cite{feidel-mertz} with a special educational concept. The buildings had been a summer recreational home, neighbored by the estate of Albert Einstein, which later was included into the school buildings.  Ernst Ising became after his dismission in 1934 a teacher and in 1937  headmaster at Caputh (see the picture at the beginning of the section). 

In the following  the political situation changed dramatically in Germany reaching its culmination in the November Pogroms. The antisemitic and racist Nuremberg Laws where enacted on September 15, 1935. The antisemitic atmosphere finally exploded in violence. Also in Caputh was Heinz Bonnem, it is  reported\cite{bonnem}: ``On the morning after the November pogroms, on November 10, 1938, the school was stormed by men from Caputh and destroyed to such an extent that on November 15, the headmaster at the time, Dr. Ising sent the message to the district school board in Potsdam, that {\it resuming classes in the foreseeable future is unthinkable}. At that time, Heinz Bonnem was a student of 4b, as can be seen from a directory from 1938." Similar can be found in the memoirs of Sophie  Friedlander and Gertrud Feiertag\cite{feidel-mertz}: ``One of the Nazis was standing outside before the door of the house. With a motionless face, he looked at his watch. As far as possible we will let everyone know, that we want to gather together on our sports field in the forest. There Dr. Ising, the headmaster, made lists and everyone, who resided in Berlin and could accommodate children, took over a group, so that everyone had a temporary home. Ernst Ising went back to Caputh  and took two pictures of his ravaged apartment. The country school home was not reopened after this destruction."

Because of these two pictures Ernst Ising was arrested by the Gestapo but was back after four hours. He could explain that he made them for the insurance company and not for propaganda.

\section{The impact and development of the Ising model until 1939 \label{impact}}

\subsection{Nordheim, Bethe and Peierls}

Friedrich Hund characterized in 1972 in his {\it Geschichte der physikalischen Begriffe} [{\it History of the physical terms}\cite{hund1972} p.364]  the understanding of ferromagnetism in the mid-30ies as follows
: ``There are still gaps between the three floors of the theory of ferromagnetism'', where  he named the three floors the qualitative quantum theory, the phenomenological - macroscopic equations, and rules for technical magnetization [material constants]. He also refers to Ising's model however erroneously  mentioned that Ising's result were equvalent to the results of Weiss' theory. This goes back to Lothar Nordheims  contribution\cite{nordheim1934} in 1934 {\it Quantentheorie des Magnetismus} [{\it Quantum Theory of Magnetism}) to Mueller-Pouillet’s {\it Lehrbuch der Physik} [{\it Textbook of Physics}]. By the way, one of the rare publications which Ising was aware of where his paper was cited. Nordheim (1899 - 1985), almost of the same age as Ising finished his thesis in 1923 by Max Born in G\"ottingen and left Germany and immigrated to the United States 1934. Ising might have known him from his time when he started his physics studies in G\"ottingen. 

Nordheim starts with a short review and a critic of the classical theory (p. 801).  He notes: (1) that there is {no explanation of the strength of the inner field by a classical interaction} (dipolar magnetic interaction is too weak); 
(2) that the theory contains inner inconsistencies, since the susceptibility should be zero and names the thesis of J. van Leeuwen. In consequence he summarizes: ``The whole complex of appearances of the magnetic properties of matter is therefore to be addressed as a pure quantum effect."
(3) Later in the paragraph on ferromagnetism he again emphasizes (p. 842) that there is no classical explanation of the inner field and cites Yakov Frenkel who made this point based on Yakov Dorfman's experiments but did not work it out\cite{ObitDorfman}. This was done by Werner Heisenberg. 
(4) He then explains Heisenberg's model and  Bloch's calculation of the magnetization at low temperature\cite{bloch1930}. An important feature of the result was that a finite limiting value of the magnetization at $T=0$  is obtained only in three dimensions. 
In this context he cites Ising's publication (p. 859) with the comment: ``Incidentally, this [referring to Bloch's calculation] is in agreement with the result of the classical theory that in a linear chain, if only the dipole interaction of the individual magnetons is taken into account, no ferromagnetism is possible [citation of Ising's paper].'' This is misleading since the Ising model is not based on the dipole interaction but assumes a short-range interaction between the magnetons. However it shows the importance of the space dimension for the Ising and Heisenberg model.
Felix Bloch\cite{bloch1932} showed for the Heisenberg model that the magnetization decreases from its maximum value as $M=M_0(1-(T/T_c)^{3/2})$. 
This decrease is a consequence of the existence of spin waves in the ferromagnetic phase of the Heisenberg model. The three-dimensionality of space turns out to be  an essential condition for the validity of this result. 

\begin{figure}
 \includegraphics[height=0.38\textwidth]{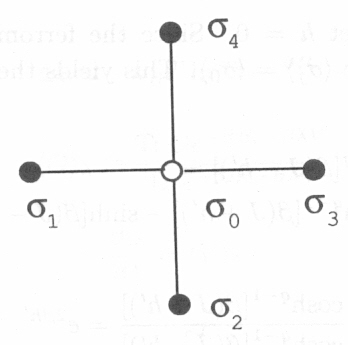}
 \hspace{0.3cm} \includegraphics[height=0.38\textwidth]{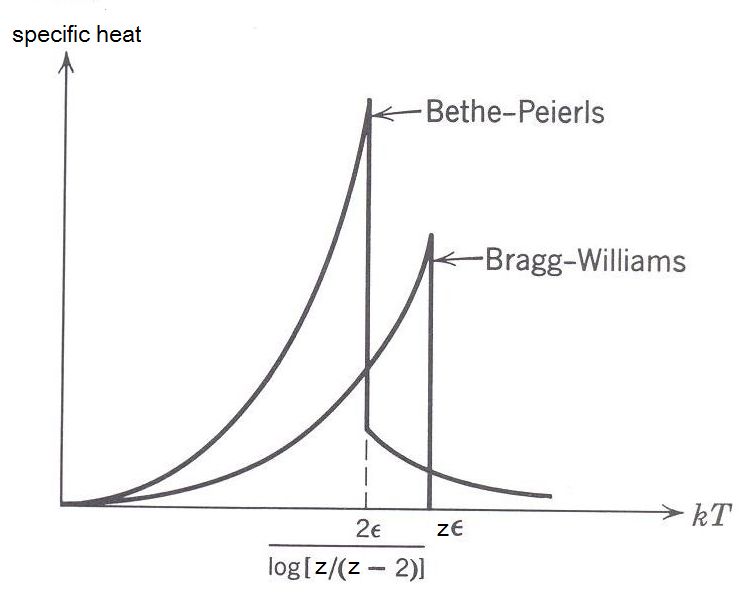}
 \caption{\label{bethe} Bethe-Peierls approximation for a two-dimensional Ising model:  (a) the neighboring spins $\sigma_1$ to $\sigma_4$ of $\sigma_0$ whose configurations are taken into account by the Bethe's approximation, 
  (b) the jumps of the specific heat in two dimensions in mean field theory (Bragg- Williams) and the Bethe-Peierls approximation. (slightly changed taken from Ref. [\refcite{huang}])} 
 \end{figure}

As already noted, Hans Bethe, who was of Jewish ancestry, was also dismissed from his position as an assistant professor at the University of T\"ubingen. He accepted a temporary lectureship in Manchester, UK, before emigrating to the US in 1935 to join the faculty of Cornell University. In Manchester Sir Lawrence Bragg suggested him\cite{mermin} a problem appearing in binary alloys in mean field theory. This approximation called for alloys -  Bragg-Williams approximation - was completely independent from the geometrical configuration of the two types of atoms in the alloy and an ordered phase existed even for the one-dimensional case.  Bethe tried to correct\cite{bethe1935} this in 1935. In order to do so he explicitly included in his mean-field calculation the configuration of the neighboring atoms (in magnetic systems this would correspond to the spin orientation, see Fig. \ref{bethe}(a)). 

In consequence the transition temperature became now dependent on the number of nearest neighbours
\begin{equation} \label{bethepeierls}
 kT_c=\frac{A}{\log(z/(z-2))} 
 \end{equation}
but the jump in the specific heat remained (see Fig. \ref{bethe}(b)). The important change was now that  the phase transition temperature in the one-dimensional case turned out to be zero. Bethe was aware of the analogy to the ferromagnet (he explained this in the introduction) but he seemed to be not aware of Ising's publication from 1925. He didn't cite Ising but remarked again at the end of his paper when he pointed to the restriction of the ordering only in two or more dimensions: ``This should be noted in connection with ferromagnetism which is restricted to {\it three} dimensions, due to a quantum mechanical effect.''

The approximation is usually called Bethe-Peierls approximation since it seems to originate from the close cooperation in Manchester between Bethe and Peierls as the acknowledgement indicates. Indeed Peierls, who decided not to return to Germany, worked on similar problems and published a paper on his own\cite{peierls1} later when he got a position at Fowler's institute in Cambridge. His acquaintance with Hans Bethe was significant\cite{shifman}, a fellow student one year his senior, knowing each other from the years in Munich in Sommerfeld's group. ``They shared an interest in and a passion for physics that resulted in a lifelong friendship that went far beyond the research-related acquaintance." In fact the formula above, Equ. (\ref{bethepeierls}), for the transition temperature has been given by Peierls. He also does not mention Ising but turned to the Ising model in the following paper\cite{peierls1936}.

The close connection of the mathematical treatment in statistical mechanics of phase transitions in alloys and  in ferromagnets led Peierls to study Ising's model\cite{peierls1936}. He considered the problems to be mathematically equivalent and cites studies of Bragg and Williams, Borelius, Fowler. Bethe and others. However Heisenberg's model was considered to be more realistic for ferromagnets not knowing which properties of the model are important for the aspects of the phase transition. For instance Peierls points to the difference between the two models with the remark that Heisenberg's model ``depends not only on the arrangement of the elementary magnets, but also on the speed with which they exchange their places." It was this difference (Peierls' interpretation of the exchange interaction) which led him to the conclusion: ``The Ising model is therefore now only of mathematical interest."

\begin{figure}\centering
 \includegraphics[height=0.35\textwidth]{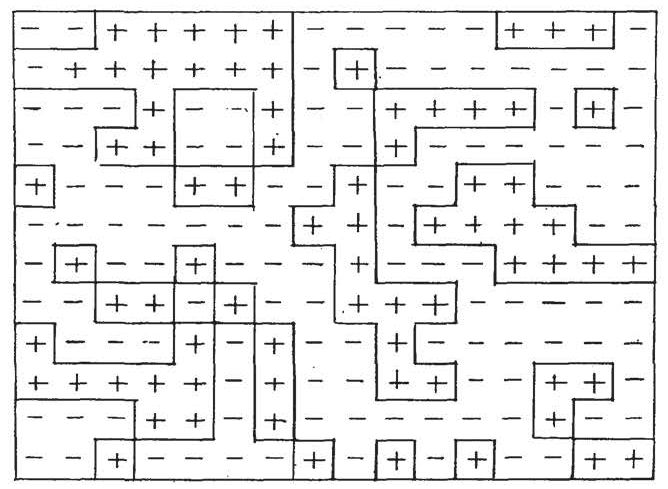}
 \caption{\label{peierls} Islands of different spin orientation for the planar Ising model from Ref. [\refcite{peierls1936}].}
 \end{figure}
Basis of his consideration was the result in the above mentioned publication\cite{peierls1} where within the Bethe approximation the nonexistence of a phase transition in one dimension was shown. ``Since, however, the problem of Ising's model in more than one dimensions has led to a good deal of controversy \ldots it may be worth while to give its solution." And indeed he could end his paper with the statement:
``Thus it follows rigorously that for sufficiently low temperatures the Ising model in two dimensions shows ferromagnetism and the same holds a fortiori also for the three-dimensional model.''
Although the argumentation turned out not to be strict the conclusion was right as it was shown in 1964 by Griffith\cite{griffith1964}. Peierls proof was also  extended\cite{bonati} to higher dimensions.
Peierl's method proved a phase transition but could not calculate the transition temperature, given later by Kramers and Wannier (see Sect. \ref{dual}).  1960 Domb\cite{dombBl} pointed out that the Bethe approximation is exact on lattices that have no closed circuits of n-n bonds [Bethe-lattice].

\subsection{The Ising model in textbooks}

\subsubsection{Van Vleck 1932}

The first citation of the Ising model in a textbook is found in Van Vleck's textbook\cite{vanvleck32} {\it The Theory of Electric and Magnetic Susceptibilies} published in 1932. Van Vleck had made his thesis in 1922 in the States supervised by  Edwin C. Kemble who was a pioneer of quantum mechanics in the States. Van Vleck is considered as a Father of Modern Magnetism and got the Nobel Price in Physics in 1977 for his fundamental theoretical investigations of the electronic structure of magnetic and disordered systems together with P. W. Anderson and N. F. Mott. 

In the preface of his book he wrote:  ``The new quantum mechanics is perhaps most noted for its triumphs in the field of spectroscopy, but its less heralded successes in the theory of electric and magnetic susceptibilities must be regarded as one of its great achievements.'' He remembers\cite{biogr77}: ``My doctor’s thesis was the calculation of the binding energy of a certain model of the helium atom, which Kemble and Niels Bohr suggested independently and practically simultaneously, with Kramers making the corresponding calculation in Copenhagen. The results did not agree with experiment for the {\it old quantum theory} was not the real thing. However, when the true quantum mechanics was discovered by Heisenberg and others in 1926, my background in the old quantum theory and its correspondence principle was a great help in learning the new mechanics, particularly the matrix form which is especially useful in the theory of magnetism." 

In chapter XII \S 77 he discusses Heisenberg's quantum mechanical theory of ferromagnetism. Taking Heisenberg's spin  Hamiltonian in the classical limit Van Vleck argues that one would yield the Weiss result and concludes erroneously [page 331 footnote 19]: ``In seeming contradiction, Ising found that classically there was no ferromagnetism regardless of the crystalline arrangement [citation of Ising's paper]. This, however, was because Ising arbitrarily took the coupling between elementary magnets to be proportional to $\mu_{x_1}\mu_{x_2}$ rather than to the complete scalar product $\mu_{1}\mu_{2}$ [$\mu$ represents the quantum mechanical spin vector]". This is different from Pauli's comment at the Solvay conference, which Van Vleck had attended. However Heisenberg's model did show a phase transition in three dimensions but Heisenberg used the Gaussian approximation (the same holds for Ising's model in mean field theory but this was never studied at this time). Niss\cite{niss1} considers this remark as so serious that the Ising model was dismissed as a realistic model in the early 1930s.

Concerning Heisenberg's model Van Vleck also mentions that a classical analysis of this model would yield the Weiss result. The reason was that Heisenberg in his paper made the Gaussian approximation as mentioned above and therefore got the mean-field result. Some pages later [on p. 344] he noted: ``Even in absence of a magnetic field, there should be a discontinuity in specific heat as one passes through the Curie point, a result  first noted by Weiss. In terms of Heisenberg theory, this is because...''. Again a result of the approximations made.

\subsubsection{Bitter 1937 \label{bitterbook}}

After his Bachelor of Science degree in 1925 at Columbia University Francis Bitter went to Europe for one year and studied in Berlin. Bitter recalled\cite{martin} hearing Max Planck's lecture on thermodynamics, attended the colloquium at which Erwin Schr\"odinger introduced wave mechanics, and teached himself electricity and magnetism from Abraham's textbook, {\it The Classical Theory of Electricity and Magnetism} (the English translation appeared later in 1937). Returning back to the States he received his PhD in 1928 from Columbia University. Rabi who visited Hamburg University for one year remembers\cite{rabiPT}: ``During the exciting period 1925 - 28 when the world of physics was reborn with the invention of quantum mechanics. It was a wonderful time to be a graduate student with a lifetime before one for research and study and the exciting task of remaking the old physics and bringing on the new.'' Bitter established The National Magnet Laboratory at MIT in Cambridge, Mass. in 1938, worked at Westinghouse and later became university professor.

\begin{figure}[h]\centering
\includegraphics[height=0.4\textwidth]{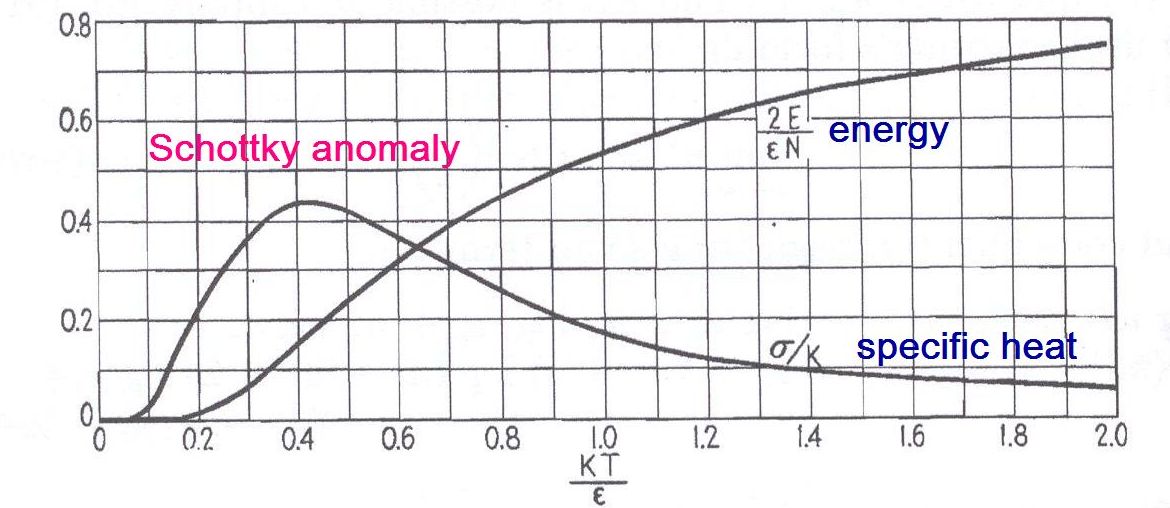}
\caption{\label{bitter} The scaled energy and specific heat against the scaled temperature for the Ising chain (adapted from Ref. [\refcite{bitter37}])}
\end{figure}

In his\cite{bitter37} {\it Introduction to Ferromagnetism}  in Chapter V Bitter presented {\it Ising's treatment} of the linear chain. He noted ``The derivation given in this section has been taken from Ising's unpublished dissertation at the University of Hamburg." And indeed he presents the complete detailed calculation as indicated in Chapter \ref{hard}. Moreover he calculated in addition the thermodynamic quantities as the energy and the specific heat $\sigma$ (see Fig. (\ref{bitter})), the last is given by (in his notation)
\[
\frac{\sigma}{k_B}=\Big[\frac{\epsilon/(2k_BT)}{\cosh(\epsilon/(2k_BT))}\Big]^2 .
\] 
The appendix contains F. Zwicky's updated article {\it Cooperative Phenomena} published already in 1933. He points to general problems in understanding phase transitions: (1) how a system of atoms can produce
{\it long-distance order}, (2) a {\it sharp transition} from one phase into another. However no conclusive answers were given and the problem remained.

\subsubsection{Becker, D\"oring 1939}

Richard Becker was born in Hamburg and got his doctoral degree in 1909 under August Weismann - a evolutionary biologist at the University of Freiburg. After hearing lectures by Arnold Sommerfeld at the Ludwig Maximilian University of Munich, Becker turned his professional interest to physics. He also studied physics under Max Born at the Georg-August University of G\"ottingen,  Becker completed his Habilitation in 1922 under Planck and became a Privatdozent at the University of Berlin. In 1926, he was appointed a full professor at Technische Hochschule Berlin  and  head of the new physics department there. In 1933 he was reassigned  to G\"ottingen as a part of the general anti-theoretical policies promoted under Deutsche Physik - his position at Technische Hochschule Berlin was eliminated. 

Werner D\"oring's thesis was supervised by Becker in Berlin and D\"oring followed him in 1939 as Privatdozent (assistant professor) at the University of G\"ottingen. After the war he returned to G\"ottingen and became in 1977 professor at the University of Hamburg.

The book\cite{becker39} {\it Ferromagnetismus (Ferromagnetism)} started from lectures given by Becker in Berlin in 1934/35. A remarkable document is the review of this book by L. F. Bates in 1946 in 
Nature\cite{bates46}. He says: ``It is very unfortune that this important work should have appeared at such an inopportune moment in the world’s history, when it will be difficult for most of the English-speaking peoples to obtain copies of it. Those of us who had the pleasure of meeting Prof. R. Becker at the 1939 Conferences on Magnetism in Bristol [The conference\cite{bristol39} {\it Internal Strains in Solids} July 11 - 13, 7 weeks before the war broke out in 1939] and in Strasbourg must sympathize with him in no small measure.'' And Kittel\cite{kittel88} wrote: ``In the decade of the 1940s physicists everywhere were educated in the principles of ferromagnetism (and learned German besides) from that excellent monograph Ferromagnetismus by R. Becker and W. D\"oring, published in 1939. It was a decade in which many physicists worked in ferromagnetism through the degaussing of ships in wartime; the development of permanent magnet materials; ...
Becker and D\"oring was an exciting and thorough book. There was a problem running through all of {\it Ferromagnetismus}, like the cliche of Hamlet without the Prince of Denmark: there was no domain theory, no
theory of the origin, size or shape of domains."

In chapter 5 the Curie-Weiss theory is explained and criticized on the basis of statistical mechanics. The authors explain the ordering through  flocking behavior of elementary magnets ({\it Schwarmbildung}) due to the short-range interaction. In this way a cloudy structure (domains) of parallel elementary magnets are created. One might guess that this modern picture (see the chapters in this volume on flocking theory) goes back to Becker's biological education in Freiburg.

\begin{figure}[h]\centering
\includegraphics[height=0.4\textwidth]{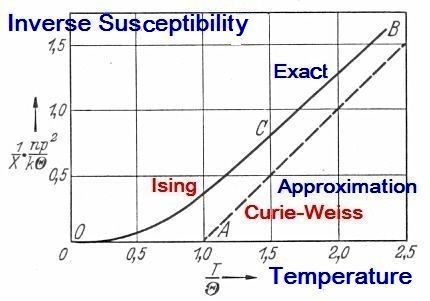}
\caption{\label{suscept}The scaled inverse susceptibility against scaled temperature for the Ising chain (adapted fromRef. [\refcite{becker39}])}
\end{figure} 
In this context also the Ising chain and its susceptibility is compared with the result of Weiss theory (see Fig. (\ref{suscept})). They consider this model as ``a first step from the Langevin case of paramagnetism of free elemental magnets in the direction of the case of ferromagnetism that actually interests us.''
The susceptibility below $\Theta$ in this approximation  is not shown. It would be a dashed line reflected on the vertical at $T=\Theta$, finite at $T=0$ and zero at $T=\Theta$ with a slope twice as steep as above $\Theta$.

\subsubsection{Extension to Condensation 1939}

An important step was made by Felix Cernuchi and Henry Eyring\cite{cernuchi39,brush83} (at Princeton University) by extending the theory of alloys to liquids by replacing one constituent of the alloy by a hole in the lattice model. This allowed them to get phases with different density. Cernuchi was a post-doc  and student of Ralph H. Fowler; Eyring had a position as an instructor at Princeton 1931 and became full professor 1938. 

Such a lattice theory was seen as an application of the Ising model to critical phenomena of liquids. In a {\it Historical Note} in the introduction to their paper they explain the steps leading from Ising's publication to their theory. As important basis for developing statistically the model they cite Bethe's and Peierls' paper discussed above. They conclude: ``the existence of two different phases at equilibrium is a direct consequence of the attractive forces between the particles.'' Their calculations are within the approximation of Bethe and Peierls and they illustrate  their result by a general phase diagram for fluid systems.

\subsection{The Van der Waals Centenary Congress in Amsterdam  November 23 - 26, 1937}

 Already in November 1937, a conference  took place in Amsterdam in honor of van der Waals (Johannes Diderik van de Waals was born hundred years before, on November 23, 1837 in Leiden). Phase transitions was a highly discussed topic. Questions like how can the gas molecules “ know” when they have to coagulate to form a liquid or solid, or could  the partition function for a finite system explain a sharp phase transition and is one partition function enough for two phases, were discussed. Kramers (chairman of one session on phase transitions) let vote the audience\cite{dresden1} with the result that no majority for one of the opinions could be found. Kramers  suggested\cite{dresden2, cohen} the use of the {\it thermodynamic limit} as an explanation of the sharp phase transition and the existence of different phases. He named ferromagnetism as an example for that behavior. 
 
 Kramers suggestion has been put on a rigorous basis later by Yang and Lee, whose theory relies on the well-known mathematical fact that a sequence of smooth functions can have a non smooth limit, provided that the convergence is not uniform. Leo Kadanoff\cite{kadanoff} explained in 2010: ``Throughout, the behavior at the phase transition is illuminated by an ``extended singularity theorem”, which says that a sharp phase transition only occurs in the presence of some sort of infinity in the statistical system. The usual infinity is in the system size. Apparently this result caused some confusion at a 1937 meeting celebrating van der Waals, since mean field theory does not respect this theorem. In contrast, renormalization theories can make use of the theorem.''

\section{Escaping to Luxembourg}
 \parbox{0.5\textwidth}{
\begin{quote}
Wer half mir\\
Wider der Titanen Uebermuth?\\
Wer rettete vom Tode mich, ...
Von Sklaverey?
\end{quote}}
 \parbox{0.55\textwidth}{
 \begin{quote}
Who helped me\\
Against the Titans' insolence?\\
Who rescued me from certain death,\\
From slavery?
\end{quote}}\\
The lines written above are from {\it Prometheus} by Johann Wolfgang Goethe  which Ernst Ising could  recite. He always had with him a little book with poems of Goethe to read and  correct if he lost some words. His son Tom Ising remembers\cite{IFKBH2017}: ``My father liked acting and had a stage in the basement where he and his friend, Heinz Wildhagen put on plays. Heinz spent his life as an actor and theater owner. Later the actor Willie Busch became my father’s dictation coach.''\vspace{0.5cm}\\
 \parbox{0.40\textwidth}{
\includegraphics[width=0.4\textwidth]{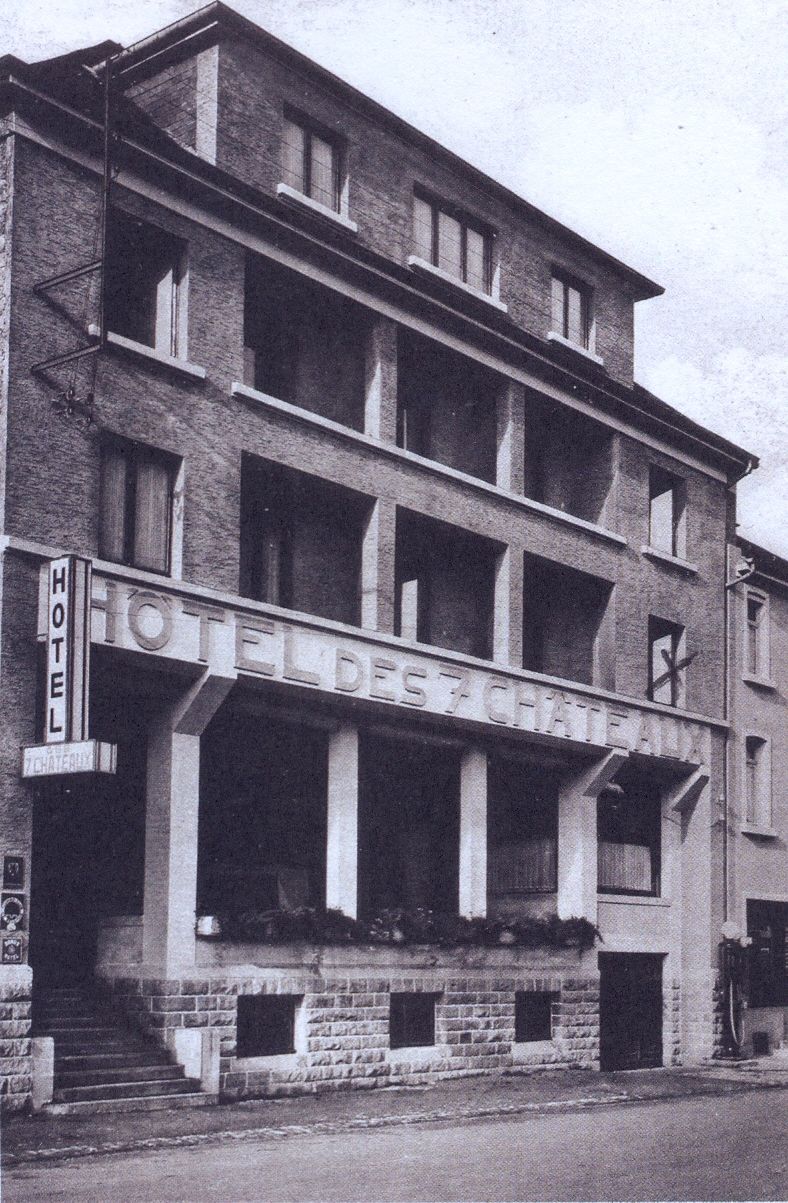}\\ The first quarter in Mersch {\tiny \copyright Collection Julie M\"uller-Barthelemy}}
\hspace{0.05\textwidth}
\parbox{0.50\textwidth}{
\begin{enumerate}
\item[1939]  January 27  Gestapo arrests Ernst Ising  for a 4hour questioning\\
May 1 arrival in Mersch\\ September 1 Begin of  World War II.\\ September 9 Birth of son Thomas
\item[1940] May 10  Occupation of Luxemburg
\item[1941] teaching at a Jewish school until October 
\end{enumerate}
} \vspace{0.5cm}

After the short arrest of Ernst Ising Johanna and Ernst  decided to ask for  emigration to the United States. Since the approval needed about two years and their situation became more and more unsafe they organized their emigration to Luxembourg for waiting to get the permission. On May 1, 1939 they arrived in a hotel in Mersch\cite{goetzinger}  where they got one of the rooms (with 19 other families in the hotel). Later, after the birth of the son Thomas on September 9, 1939, they moved to a two room flat in Berschbach/Mersch. 

On May 10, 1940 - Ernst Ising's 40th birthday - the Germans occupied Luxembourg and declared it part of the Greater German Reich. In consequence German laws now also hold in Luxembourg with severe changes for their live. On September 5, 1940, Gauleiter Gustav Simon introduced the ``Ordinance on Measures in the Field of Jewish Law", the first of a whole series of anti-Semitic ``ordinances" that strictly regulated the lives of Jews and largely restricted their freedom\cite{schoentgen}.


One month later in October 1940 there was a rumor that they had to leave either to the occupied Belgium or France or  to the unoccupied part of France. Fortunately they could stay and in December Ernst found a job as a teacher for Jewish children, which were similar to the year 1933 in Germany thrown out of public schools and collected in a separate school set up in Luxembourg-Ville.
 
But time was running out. On July 1, 1941, the U.S. State Department centralized all alien visa control in Washington, DC, so all applicants needed to be approved by a review committee in Washington, and needed to submit additional paperwork, including a second financial affidavit. At the same time, Nazi Germany ordered the United States to shut down its consular offices in all German-occupied territories. After July 1941, emigration from Nazi-occupied territory was virtually impossible\cite{holocaust}. 

Johanna wrote\cite{johannaWalk}: ``In 1941 the two-year waiting period was over, the money for the trip from the Spanish border and for the boat passage was available, almost all papers were ready.  Only one more specification [affidavit] was required of where our sponsors would lodge us in their house, how much pocket money they would give us and more such details. We wrote, but before the answer arrived the American consulates closed in June 1941 and no more visas given out! It was a terrible disappointment and meant that we would to stay in Luxembourg under the rule of the Nazis until the War was over.'' Between 1938 - 1941, US law allowed only 27,370 immigration visas per year to be issued to people born in Germany or Austria. 

\section{The international Conference on Magnetism in Strasbourg 1939}

\begin{figure}
\includegraphics[width=0.95\textwidth]{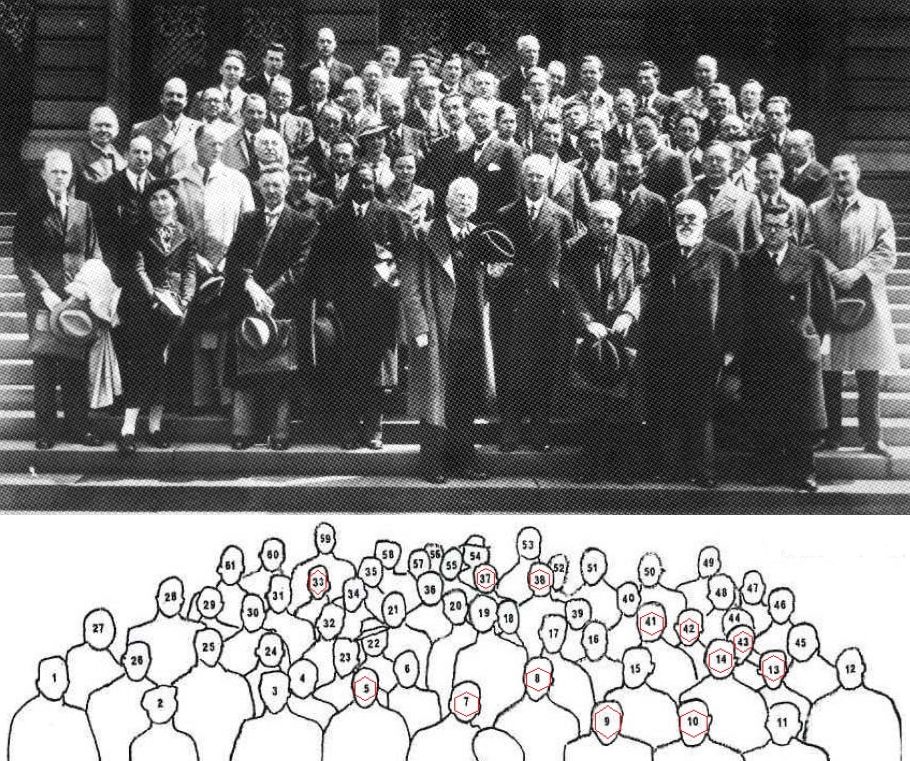}
\caption{\label{stras1939} Conference on Magnetism in Strasbourg  May 21, 1939. Speakers were  {33 Gorter}, {37 Forrer},  {38 Mott},  {41 Casimir}, {42 Foex},  {43 Neel},  {14 Kramers}, {13 Sucksmith},  {5 Krishnan}, {7 P. Weiss}, {8 Barnett}, {9 Cabrera},    {10 Bizetti}  for further identification of the participants see the reference  [\refcite{pweiss}]. \copyright AIP Emilio Segr$\grave{e}$ Visual Archives}
\end{figure}

While Ernst Ising prepared himself together with other emigrants by learning English and Spanish for the new chapter of his life, 175 kilometers south  in Strasbourg the last international conference on magnetism took place from May 21 to 25, 1939. This was after the Solvay conference in 1930, the second major international conference devoted to Magnetism and as Elliot \cite{elliott} claims: ``the fundamental ideas which underpin our understanding of magnetic phenomena today were largely in place. The Strasbourg conference of that year was able to look back also to the triumphs of the classical era.''  Indeed the development of the international  conferences on magnetism after the war showed this clearly. 

S. J. Barnett\cite{barnett1946} gave a complete report on the conference which was published  in Science as late as after the Second World War in 1946. There was only a limited number of contributions [in total 18] allowed, and attendance\cite{bates1972} was by invitation. The speakers had to submit their contributions in advance, which were then distributed before the conference to the participants. In addition the gathering was focused around six discussion sections lasting three hours each\cite{wills}. However although the proceedings were published in 1940 in France their distribution was long prevented by the German occupation.

Kramers contribution is of special importance for the Ising problem although no explicit results were presented. Therefore his talk\cite{kramers1939} {\it The Interaction Between Magnetic Atoms In A Crystal} was summarized by Weiss: ``he takes the point of view of a rigorous quantum-mechanical treatment... Mr. Kramers shows the difficulties encountered by various previous theories in demonstrating the existence of a Curie point and concludes: An exact solution that would prove strictly the existence of a transition point has not been given for any interaction model.'' 

Nevertheless some preliminary result were mentioned by Kramers, who had already started to work on ferromagnetism\cite{heller1934} in 1934 where he observed\cite{dresden1} ``that  the partition function can be obtained as the largest eigenvalue of a matrix'' and ``that the model which he constructed does not show ferromagnetism in one and two dimensions but does show it in three dimensions. This result demonstrated that the ferromagnetic phase transition is possible in a classical context and thereby settled an important issue.''  Kramers continued this investigations of the ferromagnetic transition in a paper\cite{kramers1936} published in 1936.

In his Strasbourg contribution Kramers gave a review of the present understanding of ferromagnetism in solids. Although only approximate results as in mean field approximation or in Gaussian approximation are available, he announced   some important unpublished steps for proceeding with theory and made remarkable statements concerning the understanding phase transitions in general. 

Points made by Kramers were: (1) the importance of the thermodynamic limit, (2) the description the two phases - high and low temperature phases - by one partition function, (3) the importance of the dimension of the many-body system for the existence of a phase transition, (4) the announcement of a new way of calculation of the partition function which reproduced Ising's result, and  (5) the existence of a special property relating the two magnetic phases. But no further mathematical details were given.

Considering the results in one dimension for the Ising model and the Heisenberg model he noted\cite{kramers1939} (on p. 53): 
``The difficulties that prevent the exact solution of an interaction problem for a two- or three-dimensional network are great. However, we have the impression that this solution may be found one day. We foresee that for the two- or three-dimensional network a point of transition may well exist thanks to the infinite number of ways, including one atom A can in the lattice be linked with another atom B by a continuous chain of direct neighbour-to-neighbour interactions. Indeed, the state of a solid below a transition point is a {\it long-distance order} state, which we would like to see result from an elementary law of interaction, whose direct effect would only be a {\it small distance order}.'' (see item (3) below in Sec. (\ref{ons44}) on the Onsager solution)

He further remarked\cite{kramers1939} (on p. 55): ``The tendency to ferromagnetism is greater in the model of Lenz-Ising than in that of Heisenberg. It seems likely to me that the Lenz-Ising model is already ferromagnetic for a two-dimensional network . The author [its himself] has developed a new method (not published) which, for dimension one, leads to a result identical to that of Ising. For two dimensions we find a Curie point, but unfortunately in this case it was necessary to make in the calculation use of unsure simplifying assumptions and that is why this proof of the existence of the Curie point cannot be considered as sufficient.'' D. Ter Haar writes\cite{terhaar} in {\it Masters of Modern Physics The Scientific Contributions of H. A. Kramers} (on p. 6 in footnote 8) that Dresden had informed him that according to Uhlenbeck: ``Krames had already found the transfer matrix method in 1937 and had mentioned it at the Van der Waals Conference in 1937.'' This might just refer to the method presented in the 1934 and 1936 papers of Kramers which considered the Heisenberg model rather than the Ising model. However the remark pointed to the fact that the spatial dimension is an essential parameter for the existence or non-existence of a phase transition. Indeed in 1966 N. D. Mermin and H. Wagner could  proof\cite{merminwagner66} the absence of ferromagnetism and antiferromagnetism in one- or two-dimensional isotropic Heisenberg models.

Kramers also considered the new phenomenon of antiferromagnetism found by Neel in the 30ies and speculated on the transition temperature for that phase: ``However, it is easy to see that the antiferromagnetic model of Lenz-Ising is, from a formal point of view, identical to the simplest imaginable model of the ``order-disorder" problem in alloys.'' (p. 56) Thus a kind of universality appeared in connection with the Ising model.

It was already an ambivalent  atmosphere at the conference\cite{McRae2014}:  ``The Rector of the university then closed the conference...in a final toast {\it to the honour, and preservation, of free scientific thought} as it had just been exemplified at the conference itself. It was a noble thought, grotesquely at odds with a world collapsing around him.'' Indeed not all conference participants were amused\cite{keith}. As Stoner describes in his diary entry for  May 24, 1939 the evening's conversations: ``Mrs Becker was {\it rather more obstreperous...they both [also Gerlach] want Strasbourg...a German city}''. And  Bates remembers later\cite{bates1972} ``Most of us left Strasbourg with heavy forboding.''. 

Three months later on September 1 the invasion of Poland and the Second World War started.

\section{Waiting time over- Consulate closed - Trembling in Mersch}
 \parbox{0.55\textwidth}{
\includegraphics[width=0.55\textwidth]{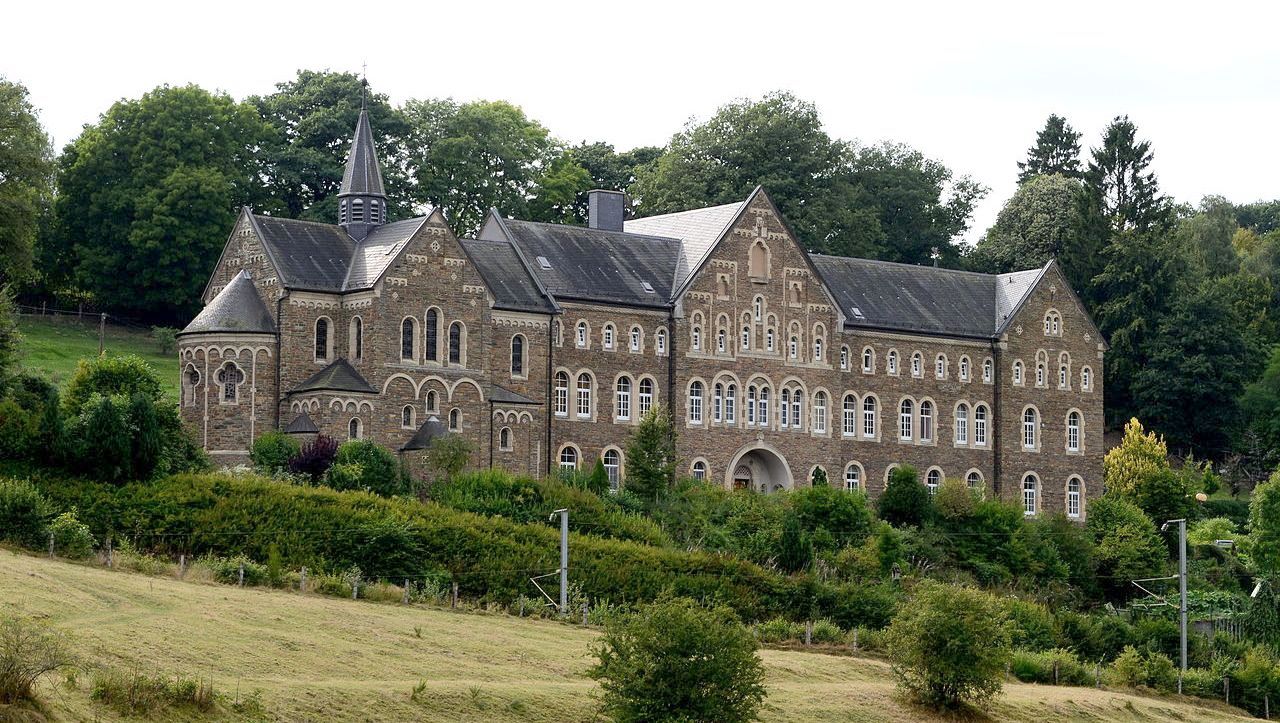}\\
Monastery Cinqfontaines (F\"unfbrunnen) the only Luxembourgian internment camp for Jews.}
\hspace{0.05\textwidth}
\parbox{0.38\textwidth}{
\begin{enumerate}
\item[1941] October 6 all Jewish families in Luxembourg - Ville were deported to concentration camps.
\item[1942] In April moving to Mersch Mozart street (see below) 
\item[1942] January 20 Wannsee Conference
\end{enumerate}
} \vspace{0.5cm}

After the annexation of Luxembourg, numerous Jews tried to emigrate\cite{schoentgen}. However, many elderly and sick people did not succeed in this. Out of concern for their well-being, the Jewish consistory initiated the plan to establish a home for the elderly in the F\"unfbrunnen monastery. In the Jewish retirement home, there was Jewish self-government under the control of the Gestapo. Limited food allocations, a ban on going out and wearing the Star of David characterized everyday life. The consistory tried to improve the situation of the residents. For example, the ban on going out was relaxed. Attempts have also been made to counteract the food shortage by growing vegetables or bartering with local farmers. 23 deaths were recorded in the camp. The reasons for this were the old age of the inmates, illnesses and the catastrophic prison conditions. Since there were neither walls nor fences, F\"unfbrunnen is considered a ``ghetto-like coercive community" and was similar to the so-called Jewish houses or Jewish dormitories. All Jews who had to give up their apartments by order of the Gestapo had to move in there. (see  Land of Memory\cite{fuenf} Kloster F\"unfbrunnen). Hugo Heuman wrote in his diary\cite{goetzinger}: ``Towards the middle of November 1941, the consistory was informed that all Jews in the country, around 350 in number, were to be concentrated in F\"unfbrunnen, for which barracks were to be built.'' 

In fact the small village F\"unfbrunnen nearby Troisvierges (Ulflingen) became generally known as the only Luxembourgian internment camp\cite{schoentgen} for Jews during the Second World War. From 1941, the enormous cloister became the collecting point for Jews.  On August 11, 1941 the first transport with about 20 Jews arrived in F\"unfbrunnen. About one month later the deportations to the concentration camps in the east began. 

Fortunately due to the German laws Jewish husbands of German wifes where exempted from the deportation and could stay with their family. A dramatic change happened after the Wannsee-Conference\cite{wannsee} on January 20, 1942. In this document (only one copy survived)  the plan for the extinction of the European Jews is documented (after a process of the decision making\cite{cgerlach}).  Until its dissolution on April 6, 1943, around 300 Jews passed the camp. Fifty of them gave Luxembourg as their place of birth. Most of them were German Jews who had emigrated to Luxembourg after 1933. 

From August 1942 Ernst worked at F\"unfbrunnen where he had to educate three of the interned children. He and his wife also had to care for the old people and help them for the deportation. ``Nothing happened to us'', Johanna wrote in her momoires, ``but the Nazi promised that those {\it lazy Jews} who lived in mixed marriages and were not affected by the deportation rulings would be put to work.''

\section{The new approach to the Ising model}

Although Kramers had announced his new method to calculate the Ising model in May 1939 the result was published about two years later due to the development of the war. The Netherlands where occupied by the Germans on May 10, 1940 leading successively to restrictions also at universities. Kramers had started a collaboration with Wannier 1939 and worked in the following years with him on the Ising problem in two dimensions. 

Gregory Wannier born in Basel made his thesis under Stueckelberg at the University of Basel in 1935 and afterwards went as exchange student for one year to the United States to Princeton. Back to Europe in the year 1938 - 1939 he worked in Bristol with Neville Mott and also with Kramers.

In the year 1941 five authors\cite{kramerswannier41a, kramerswannier41b, montroll41, montroll42, lassettrehowe41a, lassettrehowe41b} Kramers, Wannier (KW), Montroll (M), Lassettre and Howe (LH) presented a new method for calculating the partition function of the two dimensional Ising model according to the representation of the Hamiltonian given by Pauli at the Solvay conference 1930. The papers and announcements were received in the year 1941 in the following order: 
\begin{description}
\item[February 21 - 22 ] KW0: An abstract of a talk given at the 240th regular meeting of the American Physical Society at Cambridge Mass. by Wannier under the title {\it Statistics of the Two-Dimensional Ferromagnet} opens\cite{KW0} the breakthrough in the struggle to solve the Ising model. He announced: ``Rigorous statistical study of the Curie transition of ferromagnets is possible for a two-dimensional square net of spins with coupling between neighbours.'' Then he presented unpublished calculations of strips of spins by Kramers, which show that the partition function is given by the largest eigenvalue of a matrix. The existence of ferromagnetism needs at least a two-dimensional geometry. The value of the Curie temperature, where the partition function has a singularity is given and noted that the specific heat has no jump but an infinite value there.  
\item[ April 7 ] KW1: In  footnote 1 they\cite{kramerswannier41a} indicate that it was entirely written by Wannier due to the difficult communication during the wartime. In footnote 5 they state that the elegant procedure handling the transfer matrix is due to Montroll who applied it first to the theory of molecular chains. At the end of the paper they generally thank Montroll for helpful discussions. 
\item[June 6 ] M1: In chapter 1 of part II {\it General remarks} Montroll\cite{montroll41} writes: ``Wannier and Kramers have studied ferromagnetic nets in detail by a method somewhat similar to that described here. They first show that the partition function is the largest characteristic value of a set of linear difference equations that are derived by considering the change in the configuration of the lattice points of the system by the addition of one more molecule. Their final characteristic value problem [footnote 14] resembles ours rather closely. In footnote 14 the hindrance in the communication with Wannier and Kramers on this matter is noted:  ``Unfortunately the present state of world affairs has delayed publication of this work." The author is greatly indebted to Dr. Wannier for a discussion of his unpublished work; indeed to a large degree it was this discussion that inspired the writing of the present paper. Then in chapter 2 {\it Calculation of the partition function for narrow strips} he specifies  in an unnumbered footnote: ``The results of this section are not new, having been obtained by Wannier and Kramers; however, the approach, especially that in c-e, is new. [a-b presented the Ising chain and the Ising ladder c-e the extensions to 3 chains, an approximation for the ladder and the introduction of an external field for the chain, thereby reproducing Ising's result.]"
The paper ends by pointing to the future:  ``The author hopes to treat some of the other nearest neighbor problems mentioned at the beginning of this paper at some later date; and in conclusion he would like to thank Professor Onsager for his interesting discussions and for the numerous suggestions made throughout the course of this research.
\item[June 7 ] KW2: This second part\cite{kramerswannier41b} presents results from calculation mostly in approximation. 
\item[August 28 ] LH1: There is no reference\cite{lassettrehowe41a} to KW or M. Regarding the work of Lassettre and Howe Montroll remembers in the interview with Stephen Heims on  September 26, 1983: ``There is a fellow you may not think about, that's John Howe, who still exists, Lassiter [Lassettre] and Howe. I see him fairly regularly in La Jolla, and I think, see, he was a little out of the network. All the other individuals knew each other, saw each other, and sometimes it's very hard to think of who did what first, because it was such a net, whereas he was completely separate.''
\item[September 5 ] LH2: No citation\cite{lassettrehowe41b} of the other papers noted here.
\item[October 2 ] M2: In footnote 1, 2 Montroll\cite{montroll42} already cites M1 and the papers  KW1 and KW2.
\item[1943 ] Kubo: In 1943 Kubo\cite{kubo1943} published (his first paper) on this topic in Japanese. Due to the wartime it was written without knowledge of the aforementioned papers but based on Bethe's work\cite{bethe1935}. He starts to  calculate the partition function for a system of spins. He writes: ``The difficulty of this sort of approach lies in combinatory calculations of W [the Boltzmann weight], which is elementary but becomes complicated in higher approximations and is almost hopeless for rigorous treatments. ... However, one wonders if an analytic method can be devised for calculation of the asymptotic for of the expression (3) [that is the partition function in the thermodynamic limit]. The author is not able to conclude at present if this can be extended to two or three dimensions and only hopes that the present note may give some hints for such possibilities.''  He then derives the magnetization starting from the partition function for the system within an external field. Moreover Kubo proofs the non-existence of a phase transition for such systems by showing that there is no crossover of the largest eigenvalue of the `transfer matrix' with smaller eigenvalues.
\end{description}

Common to all the papers is that the occurence of a phase transition is related to the degeneracy of the two largest eigenvalues of a certain matrix of $2^n$ dimension on a $n\times n$ lattice.  In the case of the magnetic system this is the transition to the ferromagnetic phase, in the case of the alloy it is the transition to a phase of regular separation. The case of condensation of gasses is not mentioned.
This cascade of papers shows how science can progress in several ways at several places almost at the same time. But on the other hand  the success of networking is also demonstrated. It is interesting that in his interview\cite{montrollInter} Montroll  pointed, when he spoke about his connection with Kramers and the Ising model, to the importance of the\cite{montroll84} ``Vienna school of statistical mechanics''. The increase of this network to a global dimension was caused by the spreading of physicists - educated by  professors of the physical institute of the university of Vienna - all over the globe after the First and Second World War.

 Niss\cite{niss1} describes the part of the network concerning the Ising model in the States in the following way: ``Wannier taught Montroll statistical mechanics as a graduate student in chemistry at the University of Pittsburgh. Montroll subsequently spent the academic year 1940 - 1941 as a postdoctoral fellow at Yale University with Onsager where he told Onsager about his and Wannier’s ideas on the Ising problem. Kramers and Wannier were more interested than was Montroll in the physical aspects of the Lenz-Ising model. Wannier had studied the theory of transition points under Fowler in Cambridge and had attempted to apply the Bethe method to the melting process just before he began collaborating with Kramers in Utrecht in 1939.'' 
 
\subsection{Transfer Matrix for the Ising chain}

The partition function is given by the largest eigenvalue of a certain matrix, called transfer matrix. This matrix  on a $n\times n$ lattice is of  $2^n$ dimension. For a linear chain the matrix is a two dimensional matrix and the eigenvalues are easily found by solving a quadratic equation. For a ladder it is a four dimensional matrix and the exact  eigenvalues can be found only for special cases e.g.  when there is no external field. 

Following the calculation of Kramers and Wannier\cite{kramerswannier41a} the partition function $Z$ for the chain reads
\[ Z=\sum_{\sigma_1,...\sigma_n} e^{-{\mathcal  H}/k_BT}=\sum_{\sigma_1,...\sigma_n}e^{\epsilon/2k_BT\sum_i\sigma_i\sigma_{i+1}+mH/k_BT\sum_i \sigma_i} =\] \[= \sum_{\sigma_1,...\sigma_n}V(\sigma_1,\sigma_2)V(\sigma_2,\sigma_3)...V(\sigma_n,\sigma_1)= Tr V^n=\lambda_1^n+\lambda_2^n ,\]
where the Hamiltonian Equ. (\ref{Hising}) was used with $A=\epsilon/2$ and an external magnetic field $H$ is taken into account coupled to the magnetic moment $m$ of the electron spin.
 The important point was to recognize that the partition function can be written as  the product of the  transfer matrix $V$
\[V=\begin{pmatrix}e^{\beta+\alpha}& e^{-\beta}\\e^{-\beta} & e^{\beta-\alpha}\end{pmatrix} ,\]
with $\alpha=mH/k_BT$ and $\beta=\epsilon/2k_bT$. It's eigenvalues are 
\[\lambda_{1,2}=e^{\beta}\left(\cosh{\alpha}\pm\sqrt{ \sinh^2 \alpha+e^{-4\beta}}\right) .\]
Thus $Z$ 
\[Z(T,H)= e^{n\beta}\left(\cosh{\alpha}+\sqrt{ \sinh^2 \alpha+e^{-4\beta}}\right)^n ,\]
and the free energy $F$ is
\[F(T,H)=n\Big(\beta +\ln\left(\cosh{\alpha}+\sqrt{ \sinh^2 \alpha+e^{-4\beta}}\right)\Big) .\]
The magnetization $M$ is obtained by deriving the free energy
\[M=\frac{\partial F(T,H)}{\partial H}= m n\frac{\sinh\alpha}{\sqrt{\sinh^2\alpha+e^{-4\beta}}} .\]
Thus Ising's result is reproduced\cite{kramerswannier41a,montroll41,kubo1943} (apart from the difference in the temperature-dependent term due to the choice of the energy to be zero when the two spins are parallel; see also the footnote 19 in M1\cite{montroll41}).
Now in order to proceed for the two-dimensional case in the same way, one has to recognize that this is not possible including a magnetic field. However the idea is not to calculate the magnetization but instead the specific heat in zero external field. If there is a phase transition this is expected to be seen in a change of the dependence of the specific heat at the transition temperature. In Weiss' theory this was a jump at the Curie temperature.


\subsection{Duality (``the small miracle") \label{dual}}
A second important step was made by Kramers and Wannier (KW1\cite{kramerswannier41a}) only. They found a hidden symmetry of the two-dimensional Ising model in zero external field, which allowed them to proof the existence of a phase transition and to determine the transition temperature. The symmetry is based on transformation of the model interchanging links between spin and the spin variables defining a new spin lattice  (see e.g. the explanations by Krieger\cite{krieger} p. 78). 

In the case of the Ising chain in zero external magnetic field one may introduce the new spin variables by replacing\cite{mattis65, susskind} a  two linked spins $\sigma_i$ and $\sigma_{i+1}$ by a new link-spin
\[\mu_i=\sigma_i\sigma_{i+1} \,.\]
 Then the partition function can be rewritten as
\[ Z(T)=2\sum_{\mu_1,...\mu_n}e^{\epsilon/2k_BT\sum_i\mu_i} ,\]
where the factor 2 appears because of the two possibilities for the boundary spin (e.g. the first spin). This leads to the partition function of independent link-spins and the result is
\[Z(T)=2[2\cosh(\epsilon/(2k_BT))]^{n-1} .\]
Thus the original spin model of interacting chain spins has been transformed into a dual  model of independent link-spins each in two possible energy states - the model published in 1922 by Schottky.  This is the reason for the result obtained for the specific heat by Bitter in 1937 (see Sect. \ref{bitterbook}, also Bitter was obviously unaware of Schottky's paper) for the Ising chain to show the anomaly already calculated by Schottky in 1922 (see Sect. \ref{schottky}). Unfortunately neither Lenz nor Ising were interested in the specific heat, which could have been calculated using Ising's method.  

This idea of duality then has been  applied by Kramers and Wannier to the two-dimensional model. It turns out  that the dual model is the same as the original model, thus the two-dimensional Ising model is {\it self-dual} (see e.g. the explanations by Krieger\cite{krieger} p. 78).
The original and the dual lattice are square lattices, where in the dual lattice spins are in the center of the squares of the original lattice.
With the definitions
\[  K=\frac{\epsilon}{2k_BT}   \qquad \text{and} \qquad  \sinh(2K^*)=\frac{1}{\sinh(2K)}\]
it can be shown that for the partition function the relation 
\[ Z_n(K^*)=2(\sinh(2K))^nZ_n(K) ,\]
holds where $K^*\le K_c$ and $K\ge K_c$
This self duality relates the low-temperature phase region to the high-temperature region. Since they coincide at the transition temperature the equation  for $T_c$ can be formulated as
\[\sinh (2K_c)= 1 , \qquad k_BT_c=2.27\epsilon .\] 

Applying these ideas to the three-dimensional Ising model shows that it is not self dual and therefore no equation for $T_c$ results.

In the second paper of Kramers and Wannier (KW2\cite{kramerswannier41b}) it was shown that the specific heat has to be infinite at the Curie point. No further information could be given, but it was proven that the Curie-Weiss theory is not valid.

\section{D-Day and rescue}

 \parbox{0.5\textwidth}{
\begin{quote}
Ich dich ehren? Wof\"ur?\\
Hast du die Schmerzen gelindert\\
Je des Beladenen?\\
Hast du die Thr\"anen gestillet\\
Je des Ge\"angsteten?\\
Hat nicht mich zum Manne geschmiedet\\
Die allm\"achtige Zeit\\
Und das ewige Schicksal,\\
Meine Herrn und deine?
\end{quote}} 
 \parbox{0.5\textwidth}{
\begin{quote}
I honour thee, and why?\\
Hast thou e'er lightened the sorrows\\
Of the heavy laden?\\
Hast thou e'er dried up the tears\\
Of the anguish-stricken?\\
Was I not fashioned to be a man\\
By omnipotent Time,\\
And by eternal Fate,\\
Masters of me and thee?
\end{quote}}\\
From {\it Prometheus} by Johann Wolfgang Goethe\vspace{0.5cm}\\ 

\hspace{-0.5cm}  \parbox{0.45\textwidth}{
\includegraphics[width=0.45\textwidth]{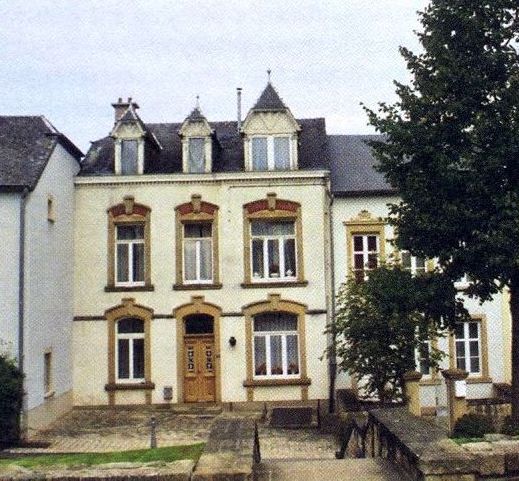}\\ Mersch Mozart street \\{\tiny\copyright Exhibition\cite{goetzinger} Mersch 2007}}
\hspace{0.05\textwidth}
\parbox{0.45\textwidth}{
\begin{enumerate}
\item[1943] May 5  unfree labor at the Maginot Line 
\item[1944] D-Day Juni 6 - September 1, no railway service, Germans flee
\item[1944]  September 10 Liberation by the Allies
\item[1947] April 9    left Rotterdam on board the freighter Lipscomb Lykes 
\item[1947] April 21  arrival in New York 
\end{enumerate}
}\vspace{0.5cm}

Ernst Ising' duty in F\"unfbrunnen ended on  April 6, 1943 when the last train with Jews left Luxembourg without him. There is a statement in Johannas memoirs which could bee seen in correspondence to Goethe's Prometheus when she says\cite{johannaWalk}: ``And I was thankful that Ernest was not among them, but instead, after having helped to make their trip as comfortable as possible, could return to wife and child and that we could face for future, uncertain as it was, together, side by side, encouraging and supporting each other. I was grateful, but to whom? To God? I would have felt conceited to think that God had selected me to be spared the horrors of deportation or separation while He allowed those 80 down there to be carried to their almost certain death.'' 
 
As it was promised by the Nazis Ising was forced to work since May 5, 1943 in Diedenhofen (south of Mersch) at the Maginot line.  At 4 o'clock in the morning he left for Diedenhofen and returned at 10 o'clock in the evening to Mersch. They had to dismantle and load railroad tracks so they could be brought to the Eastern Front and relocated there to aid the German advance. 

New hope arose when on June 6, 1944, the D-Day, the allied forces landed in the Normandy 18 years after the day Johanna met Ernst. Johanna started to note their situation in a diary (included in her memoirs\cite{johannaWalk}) starting from August 26, 1944 since she was sure that it would not fall into the hands of the Nazis.  Due to the development on the new front in France on   August 31, 1944 the work on the Maginot line was stopped.  No trains were running the next day, the Germans started to fly east and finally the Isings were freed on September 10, 1944.

The war was not over, but their lives were rescued. Ernst Ising had to look for jobs earning money. He accepted a series of odd jobs: bookkeeping, darkroom work,  and even reconstruction work as hard as he had to do at the Maginot line. In November 1946 the Isings could present themselves at the American Consulate in Amsterdam and finally in April 1947 they were able to leave for the United States!\vspace{0.5cm}

\hspace{-0.5cm}  \parbox{0.45\textwidth}{\centering
 \includegraphics[width=0.44\textwidth]{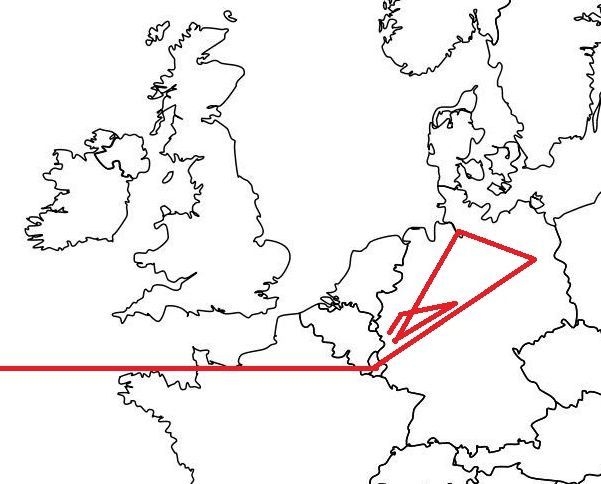}\\Path of life for Ern(e)st Ising}
\hspace{0.1\textwidth}
\parbox{0.35\textwidth}{\begin{itemize}
\item Cologne
\item Bochum
\item G\"ottingen
\item Bonn
\item Hamburg
\item Berlin
\item Mersch   
\item United States
\end{itemize}
}

\section{On the steps to the solution}

After Lars Onsager finished his studies in Trondheim in 1925 he worked on the theory of electrolytes. His first publication was a revision of the  Debye-H\"uckel theory, which enabled him to get an assistant position at the ETH Z\"urich at Debye's institute for 1927 to 1928. There he got to know the German Hans Falkenhagen who worked also at the institute. Both young scientists laid the basis for their future career there.
Onsager went back to the States and after a semester at John Hopkins University Baltimore found a position at Brown University  Providence until 1933. Then due to economic depression the university could not keep his position he had to look for a new one. He used this situation for a visit in Europe to meet Falkenhagen who had in this year a position\cite{scheffcyk} as extraordinarius professor at the university of Cologne. Falkenhagen told W. Ebeling\cite{ebeling} how the visit happened: ``The scientific discussion was restricted to the short way from the railway station to our house, since after I had introduced Onsager to my wife and her sister Gretl Arledter who stayed with us, Onsager was lost for science. He took me aside and asked wether my sister-in-law was still free.'' Margarethe Arledter - born in Maribor Slovenia former Austria - was the daughter\cite{nobelOns} of Fritz Arledter a well-known pioneer in the art of paper making in Cologne.  After a few days on September 7, 1933 Gretl and Lars got married and  returned to the States. There he got a position at the Yale University and contact to Montroll. 

 American universities during war were less busy, many students were off fighting, and scientists were doing military research. In 1942 the United States joined the anti-Hitler coalition. Many physicist were doing military research,  but as Hemmer\cite{hemmer} noted there were exceptions:``As an alien with an Austrian wife, Onsager was not part of this (he became US citizen only in 1945) and consequently had time to prepare the next jewel in the crown; the solution of the two dimensional Ising model." 
 
 The first announcement of the existence of the solution was in a discussion remark following a paper by Gregory Wannier at a meeting of the New York Academy of Sciences on  February 18, 1942 and presented on October 24, 1942 at the Meeting of the New England Section held at Hartford, Connecticut. Onsager contributed two papers, one of them - with the title {\it Crystal Statistics} - treated the exact solution of the Ising model in two dimensions. The three-sentence abstract\cite{onsager42,hemmer2} stated:\\ 
``The partition function for the Ising model of a two-dimensional {\it ferromagnet} has been evaluated in closed form. The results of Kramers and Wannier concerning the {\it Curie point} $T_c$ have been confirmed, including their conjecture that the maximum of the specific heat varies linearly with the logarithm of the size of the crystal. For an infinite crystal, the specific heat near $T=T_c$ is proportional to $ -\log |T-T_c|$."

\subsection{Onsager's solution 1944 (``the big miracle") \label{ons44}}

On October 4, 1943 (published February  13, 1944) Lars Onsager finally sent the paper with his solution to Physical Review\cite{onsager44} whose abstract now started with the sentence: ``The partition function of a two-dimensional {\it ferromagnetic} with scalar {\it spins} (Ising model) is computed rigorously for the case of vanishing field." In order to come to a solution (finding the largest eigenvalue) it was necessary to understand the mathematical properties of the transfer matrices. Onsager replied to a question how he found these properties, that he had a lot of time during the war, so he began to  diagonalize the transfer matrix (a real tour de force as C. N . Yang said). He started with an Ising ladder. Ising looked also on the ladder in his thesis\cite{thesis} in his attempt to get the magnetization in higher spatial dimensions but included an external magnetic field which makes the problem exactly unsolvable. Onsager then added a third chain and so on up to six chains. The transfer matrix grows in this way from a $4 \times 4$ matrix to a $64 \times 64$ matrix and finding that all the eigenvalues were of a special form gave him the indication that the algebra of the problem was a product algebra. Onsager could reduce the problem from studying instead of $2^n\times2^n$ matrices $(2n)\times(2n)$ matrices and then find the eigenvalues for general $n$. Therefrom he could calculate the partition function in the thermodynamical limit

\begin{equation}
 Z=\ln(2 \cosh(2\beta J))+\frac{1}{2\pi}\int_0^\pi d\phi\ln\frac{1}{2}(1+\sqrt{1-\kappa^2\sin^2\phi}) ,
\end{equation}
with $\beta=1/(k_BT)$ and  $\kappa=2\sinh(2\beta J)/\cosh^2(2\beta J)$.  

\begin{figure}[h]\centering
\includegraphics[height=0.5\textwidth]{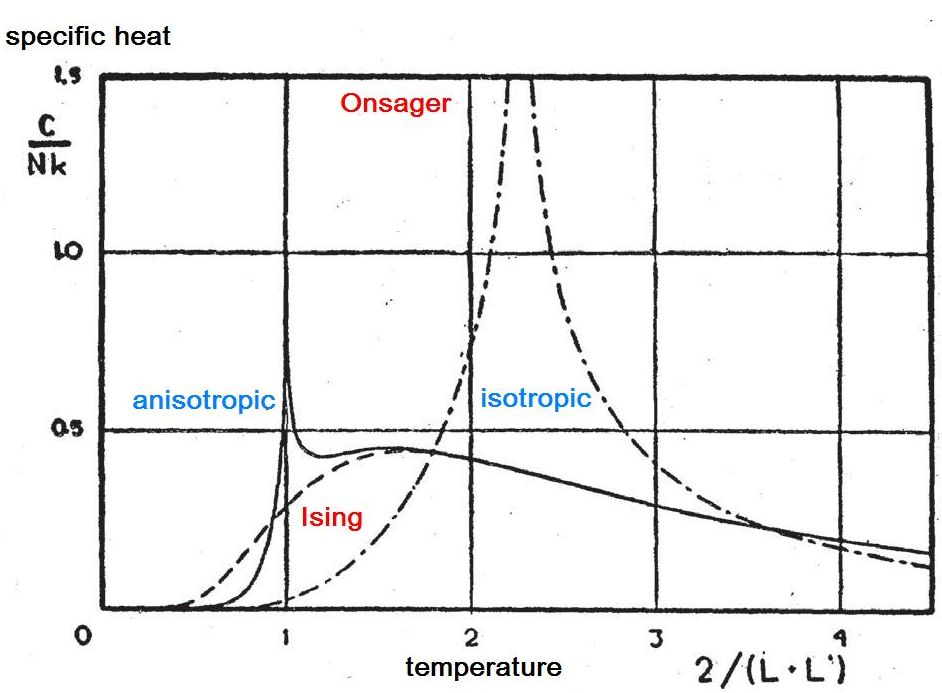}
\caption{\label{Ocross}The change of the scaled specific heat for the isotropic to the strongly anisotropic  Ising plane. In the limit of one interaction going to zero the result for the Ising chain is obtained. (adapted from Refr. [\refcite{onsager44}])}
\end{figure} 

From the solution he could calculate the specific heat\cite{onsager44,mccoyWu} showing its logarithmic divergence at the critical temperature different from the jump predicted by the mean-field theory 
\[\frac{C}{Nk_B}=\frac{2}{\pi}\big(\beta J \coth( 2\beta J\big)^2
\big[2K(\kappa)-2E(\kappa)-(1-\kappa')(\frac{\pi}{2}+\kappa'K(\kappa))\big] ,\]
with $K(\kappa)$ and $E(\kappa)$ a first and second class elliptic integral respectively and $\kappa'^2=1-\kappa^2$.

The transition temperature is given by the temperature where the specific heat diverges logarithmically as conjectured by Kramers and Wannier
\begin{equation}
\tanh\frac{2J}{k_BT_c}=\frac{1}{\sqrt{2}} .
\end{equation}

In fact Onsager solution was found for the anisotropic case with interaction strength different in the $x$ and $y$ direction of the plane. This allowed him to consider the crossover from the planar model to the Ising chain (see Fig. \ref{Ocross}). One may have speculated that the peak in the specific heat of the linear chain is in some way related to the existence of a phase transition in two dimensions. The anisotropic two-dimensional solution shows that this is not the case. The position of the phase transition temperature goes to zero whereas the peak remains in the unordered phase indicating a Schottky anomaly\cite{schottky1922} of two-level system (see Fig (\ref{anomaly}) and Sect. \ref{dual}). In fact the peak is related to the `two-valuedness' of the energy levels.

This exact solution is a cornerstone in statistical physics and it established the value of the ``simple" Ising model. It opened a new quite general task: to understand critical phenomena beyond the mean-field theory.
Onsager's solution had a much broader effect than demonstrating a ferromagnetic phase in Ising's simple model. It showed how the physical  phenomenon near a phase transition are described by our mathematical tools and what kind of features are expected (see for example the textbook of Goldenfeld\cite{goldenfeld}).

(1) It answered the question stated by Krames earlier at the Van der Waals Centennial Conference in 1937, if statistical mechanics can calculate physical quantities in different phases by {\it one partition function}. This problem lasted even longer and was taken up by Kramers also at the Florence Conference On Statistical Mechanics in 1949, the first STATPHYS\cite{hoddesonP} after World War II. Montroll\cite{montrollInter} explains: ``the big question was whether it was possible from the partition function to discuss both the ordered and disordered phase, because of an example, if one takes liquids and gases one makes one model for the solid, for solids and liquids, then you make another model for the liquid, and in each one, the partition function is calculated separately. So the basic question was, could you make a model that would have the full phase
transition and have both the ordered and disordered phase in it, and that was what Kramers was looking for with the Ising problem."

(2) Another problem was the behavior around the phase transition and the relation performing the {\it thermodynamic limit}. In mean field theory one gets two phases and critical behavior without this limit.

(3) It also answered the question how {\it long-range order} is macroscopically  established in this limit although only {\it short-range order} (nearest neighbor interaction) between the constituents of the system is present on the microscopic level. How difficult the understanding of long-range ordering was is reported in 1941 by van der Waerden\cite{waerden} in his paper where he showed that the calculation of the
Ising partition function is equivalent to the problem of counting closed polygons on the square
lattice.: ``In a recent conversation in G\"ottingen Becker doubted the strict validity of the long range. He thought that there could indeed be areas of a certain extent in which an order of the kind described prevails, but it would be difficult to imagine that the occupation of one atomic site had a noticeable influence on the occupation of another place, 10$^ 8$ atoms away.''  And even 1944 H. C. Temperley\cite{temperley44} stated: ``The assumption, implicit in certain theories, that nearest neighbor interactions are by themselves capable of leading to co-operative effects appears to be mistaken." 

(4) Variants of the non-isotropic Ising model demonstrated that the genuine critical behavior is independent of such features. This led to defining so called {\it universality classes} which show identical critical behavior although on microscopic scale they are quite different.
Examples falling into the Ising model class are: uniaxial ferromagnets or antiferromagnets, fluid systems (gas-liquid critical point), mixtures (demixing critical points), metallic alloys. The universality classed are defined by general properties like dimension of the system (e.g. one, two or three...), kind of ordering quantity - the order parameter - (e.g. a scalar, a vector or a tensor) or the symmetry of the interaction terms of the order parameter.

(5) In the thermodynamic limit when the long-range order is reached {\it singular behavior} appears in physical quantities. This turned later out also for dynamical quantities like transport coefficients when the critical behavior of dynamical systems was studied\cite{folk07}. 

(6) Onsager's solution also confirmed the conjectures of Kramers and Wannier for the specific heat of the planar Ising model with a finite number of spins. Calculation for finite systems and the dependence of physical quantities on $n$ were important for calculations of physical systems by computer simulations.  It led to what is known as {\it finite size scaling} and opened a new field for comparing physical phenomena with theoretical models in addition to experiments\cite{janke12}.

\section{Onsager  announced in 1948 and Yang published in 1952 the solution for the magnetization}

\subsection{Reactions on Onsager's solution}

Pauli\cite{pauliOns} judging the Worl War II period in physics wrote in a letter to Casimir: ``nothing much of interest has happened except for Onsager’s exact solution of the Two-Dimensional Ising Model.'' Indeed there was a big impact in the physics community increasing the interest in the Ising model. Montroll\cite{montrollInter} reports that Pauli together with someone else tried to find new results in the two- and three-dimensional case but didn't get anything out of it. In his interview Montroll also remembers:
 ``\ldots once the Onsager-Kauffman papers came  out, then it looked as though one could do much more, and in many places, the three dimensional Ising model was assigned to people as PhD theses.\dots All over the world people were told to look at the three dimensional Ising problem by the Onsager method, which it turns out of course didn't [work].'' Some of them who failed in three dimensions solved the model for other geometries like hexagonal or triangular lattices.

\subsection{Working on the solution}

\begin{figure}\centering
\includegraphics[width=0.4\textwidth]{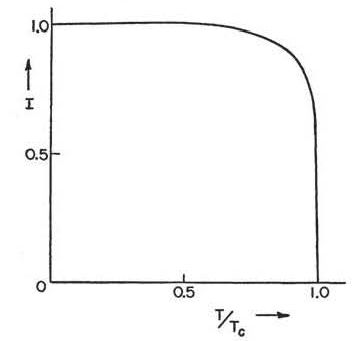}
\caption{\label{yangmag}Spontaneous magnetization of the two-dimensional Ising model as function of scaled temperature (from\cite{yangcomm1}).}
\end{figure}

Although Onsager had proven that the Ising model has a phase transition in two dimensions the calculation of the magnetization - Ising's original task - was not performed. However Onsager surprised the physics community when he\cite{longuet91}  ``silenced a conference at Cornell [in August 1948, about 800 mile west from Peoria where Ising teached at Bradley University] by writing on the blackboard an exact formula for the spontaneous magnetization.'' One year later Onsager used the first STATPHYS 1949 in Florence to announce\cite{hemmer2} again the result of a calculation together with Kaufman for the increase of the magnetization in the ordered phase. It showed definitely a deviation from mean field theory and led in the thermodynamic limit to $M\sim (T_c-T)^{1/8}$ (see  the result published by Yang in Fig. \ref{yangmag}).  

Onsager had three derivations\cite{montrollInter,baxter1,baxter2} none of them were ever written for publication because he thought they were too complex and Onsager wanted to have them simplified before he published them, so there's no publication by him on how he came to his result. 


It lasted almost 30 years after Ising's unsuccessful extensions  to get the desired explicit calculation from the publication\cite{yangcomm1} in 1952 by C. N. Yang. He had heard about the solution of the Ising model from his professor J. S. Wang at the Tsinghua National University as his Master's student in 1944 - 45. After finishing his studies he left China and emigrated to the United States. There he entered the University of Chicago 1946 and received his PHD. in 1948.  

Yang concentrates in the short abstract of his paper to the result and its physical basis: ``The spontaneous magnetization of a two-dimensional Ising model is calculated exactly. The result also gives the long-range order in the lattice."
A long discussed mystery was how order can emerge from short-range interaction of the ordering units and how it is related to the emergence of infinite ranged correlation. 

Even for Yang after all the struggles of the predecessors it was an extraordinary ambitious task as he reports \cite{yangcomm2}: ``Starting in early 1951, I worked on the Ising model problem intensively. \ldots I had been familiar with it because of my master's thesis. L. Onsager produced, quite unexpectedly, an exact evaluation of the partition function of the model in two dimensions. It was a real tour de force. I had studied his paper in Chicago in the spring of 1947, but did not understand the method, which was very, very complicated, with many algebraic somersaults. \dots But I did not drop the Ising model. I kept thinking about it, and realized that Onsager and Kaufman [1949] had obtained much more information than just the partition function \ldots Their method in fact gave information about {\it all} eigenvalues and eigenvectors. Proceeding in this direction, I arrived in January 1951, at the conclusion that the spontaneous magnetization is dependent on an off-diagonal matrix element between the two eigenvectors with the largest eigenvalues \ldots I should be able to evaluate this matrix element.   

I was thus led to a long calculation, the longest in my career. Full of local, tactical ticks, the calculation proceeded by twists and turns. There were many obstructions. \ldots I almost gave up \ldots Finally, \dots all the pieces suddenly fitted together, producing miraculous cancellations, and I was staring at the amazingly simple result \ldots "
Yang did not solve the problem by calculating the partition function in a magnetic field. He considered the spin spin correlation function in the limit of infinite separation of the spins to get the magnetization\cite{yangcomm2}.
This became possible by the work of Bruria Kaufman\cite{kaufman49} in 1949. 

 The title of Kaufman's paper {\it Crystal Statistics. II Partition function Evaluated by Spinor Analysis} indicates it as a reformulation of the first part written by Onsager using now spinors [reflecting the property of the spin under a rotation by 360° to change sign].  This reformulation simplified but also extended the results of Onsager by finding the eigenvectors of the transfer matrix. In a third part\cite{kaufOns49}  written by Kaufman and Onsager they went on to calculate some of the two-point correlation functions.


Even in 1964 Ising's task was not completely solved\cite{schultz64}: ``In contrast to the free energy, the spontaneous magnetization of the Ising model on a square lattice, correctly defined, has never been solved with complete mathematical rigor. Starting from the only sensible definition of the spontaneous magnetization, the methods of Yang and of Montroll, Potts, and Ward are each forced to make an assumption that has not been rigorously justified." 
A proof\cite{benettin} that the magnetization $M$ calculated from the $H=0$ correlation function is the same as calculated from the derivative at $H=0$ of the free energy with $H\neq 0$ was given in 1973. It is based on the duality transformation:
\[\lim_{x\to\infty} \sqrt{<\sigma_0\sigma_x>}=\partial F(T,H)/\partial H=M .\]
A discussion of the mathematical correct definitions of the magnetization and its equivalence can be found in Ref. [\refcite{krieger}] in chapter 5. The problematic is connected to the thermodynamic limit, the boundary conditions and how the external field is set to zero if the magnetization is calculated by different definitions. 
 
Barry McCoy was cited in the Introduction and one may cite him (see Ref. [\refcite{mccoyWu}], p. 248) again at this  point where both the specific heat $C$ and the magnetization $M$ are known:  ``\ldots it would be most useful if a proof could be constructed which would establish the identity of these two temperatures [of the $T_c$ of the specific heat and of the magnetization] without first having to compute explicitly both $C$ and $M$.\ldots Unfortunately no such proof has yet been found. Therefore, we have no fundamental understanding of how, or even if, the singularity in the specific heat is related to the vanishing of the magnetization.''

\section{After the rescue}\vspace{-0.5cm}

 \parbox{0.55\textwidth}{
\begin{quote}
W\"ahntest du etwa,\\
Ich sollte das Leben hassen,\\
In W\"usten fliehen,\\
Weil nicht alle\\
Bl\"uthentr\"aume reiften?\vspace{0.2cm}\\ 

Hier sitz’ ich, forme Menschen\\
Nach meinem Bilde,\\
Ein Geschlecht, das mir gleich sey,\\
Zu leiden, zu weinen,\\
Zu genie{\ss}en und zu freuen sich,\\
Und dein nicht zu achten,\\
Wie ich! 
\end{quote}}
 \parbox{0.5\textwidth}{
 \begin{quote}
Didst thou e'er fancy\\
That life I should learn to hate,\\
And fly to deserts,\\
Because not all\\
My blossoming dreams grew ripe?\\ 

Here sit I, forming mortals\\
After my image;\\
A race resembling me,\\
To suffer, to weep,\\
To enjoy, to be glad,\\
And thee to scorn,\\
As I!
\end{quote}}\\
From {\it Prometheus} by Johann Wolfgang Goethe 
\vspace{0.3cm}

\hspace{-0.5cm} \parbox{0.48\textwidth}{
\includegraphics[width=0.47\textwidth]{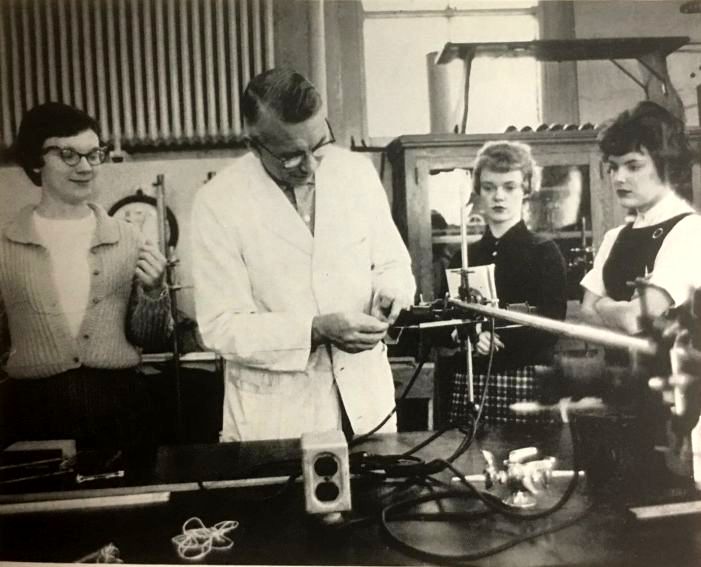} Ernest teaching in Peoria \\
{\tiny\copyright Tom Ising}}
\hspace{0.02\textwidth}
\parbox{0.45\textwidth}{
\begin{enumerate}
\item[1948] Teacher at the States Tea\-cher's College Minot (ND) 
\item[1948] Professor in Physics at the Bradley University Peoria (IL)
\item[1953] Ernst becomes Ernest and Johanna became Jane as citizens of the United States
\item[1968] Honory doctorate of Bradley University
\item[1998]  Ernest passed away on May 11 one day after his 98th birthday
\end{enumerate}
 }\vspace{0.5cm}
 
 After 12 days on the Atlantic on April 21, 1947 the Isings reached New York. On a physicists' convention in New York Ernst applied for several job offers and finally accepted a position at the Teacher's College in Minot N.D.
 They left New York on September 5 to Minot where Ernst teached physics again since 1933. Meanwhile new offers for position from the Teacher's Agency came in and Ernst took one for the Bradley University in Peoria, Illinois. Jane Ising remembered in her memoirs\cite{johannaWalk} (p. 73): ``In 1949 {\it Ernest} found out from science  magazines that his doctor thesis from 1924 on ferromagnetism has become known and that the model he had developed and calculated was called the ISING - MODEL. A considerable literature appeared about it and a two-dimensional Ising Model was developed by {\it Prof. Onsage[r]}, his own model being one-dimensional."

Ising retired when he was 76 years old due to a new regulation at the university. He was awarded in 1968 by a honorary Doctor's degree from Bradley University, in 1971 he was named {\it outstanding teacher of the year}, and in 1974 Hamburg University honored him by renewing his Doctor's degree according to the 50 years anniversary.

On May 11, 1998 one day after his birthday 98 years old he died\cite{stutz}. In 2019 his biography was included in the book\cite{markus} {\it 222 Juden Ver\"andern die Welt [222 Jews Change the World]}  describing the contribution of Jewish culture to the European and American civilization.

\subsection{Experimental proof and understanding of the rational value}

Because of their simplicity, theoretical models are often viewed as  mathematical models only not realized in nature. 
However it turns out that under certain conditions it is possible to construct real systems for which are those models apply. This happened also for Ising's model. As it was already known that the Ising model describes also the phase behavior for two dimensional fluids, the behavior of the order parameter (the density difference) has been measured\cite{kim} in the year 1984. The measurement of the corresponding magnetization in an ferromagnetic system\cite{back} followed later in 1995. The first system consisted of a sub-monolayer of CH$_4$ molecules adsorbed on graphite,  the second one measured epitaxial Fe films on top of a nonmagnetic single-crystal W surface. Both experiments were in agreement with the exact value $1/8$ of the critical exponent of the order parameter within the experimental accuracy corroborating the model and universality.

Another miracle of the model is the rational (or {\it Biblical}\cite{fisher83}) value of the critical exponent [1/8] of the magnetization.  It was solved when Alexander M. Polyakov\cite{polyakov70} suggested in 1970 an so far uncovered symmetry at the critical point -  the conformal symmetry\cite{poland16}. This symmetry emerges at the critical point where fluctuations of any scale are present (critical opalescence seen in liquids by light scattering experiments, in magnets by neutron scattering) and the system has lost its characteristic length scale. But conformal invariance goes further (see Ref. [\refcite{christehenkel}], p. ii): "While scale invariance alone is capable of casting systems into universality classes only dependent on a few selected properties like the global symmetry, the dimension of the space and of the number of components of the order parameter, two-dimensional conformal invariance yields a classification of the critical point partition functions and thereby furnishes exact values of the critical exponents. Furthermore, the critical multipoint correlation functions of the local variables of the system can be determined exactly.'' This symmetry made possible the integrability of the Ising model at $T_C$ within an external field and explains the rational exponent found in the Onsager solution.

\section{Conclusion}

The story of the struggle to solve the Ising model shows especially the complex interrelation of personal fate and scientific connections. Both might be strongly influenced by circumstances outside of the physics community. On one  hand it shows how the understanding of phenomena depends on the mathematical knowledge to treat the physical models, but also the possibilities to perform experimental verification of the theoretical models. Especially very simple models are often difficult to verify in their importance for physics since  at the time of their emergence the technological conditions to perform suitable experiments are not present. 
On the other hand the formulation of simple models and their solution are often underestimated and judged as a pure mathematical curiosity. But contrary also the Ising model  is a further example\cite{flexner} of the {\it Usefullness of Useless Knowledge}.

\section*{Acknowledgement}

 The author thanks Tom Ising for sending me the Bradley material including some personal pictures and  Sigismund Kobe for informations about Ising;  Bertrand Berche, Yurij Holovatch, and Ralph Kenna  for the longstanding cooperation on the topics treated in this paper. An additional thank goes to the Editor of this series for correction and suggestions concerning the manuscript.

The author also highly acknowledge the support by the Austrian Agency for International
Cooperation in Education and Research (OeAD) and the Ministry of Education and
Science of Ukraine via bilateral Austro-Ukrainian grant number UA 09/2020.

\section*{Note added in proof}
After this chapter was sent to the Editor  Russia  invaded Ukraine on  February 24, 2022 after having staged several false statements on the Ukrainian Government as a pretext to initiate the invasion. The Russian President Putin claimed his goal was to protect people subjected to bullying and genocide and aim for the "demilitarization and de-Nazification" of Ukraine. There has been no genocide in Ukraine: it is a vibrant democracy, led by a president who is Jewish.

This violation of the Law of Nations was explained by President Putin already a year ago in a pseudohistorical publication by denying the Ukrainian national identity and considering as Russians. The minister of Foreign Affairs of Luxemburg Jean Asselborn said: {\it February 24, 2022 is to Ukraine what May 10, 1940 was to Luxembourg}.
 
In Russia on March 4 2022 the  Russian Union of Rectors formulated a common statement in support of the ongoing military attack of a free country. This should be compared to the Vow of Allegiance of the Professors of the German Universities and High-Schools to Adolf Hitler and the National Socialistic State of November 11 1933.
 In Russia one reads: {\it It is very important these days to support our country, our army, which defends our security, to support our President, who, perhaps, made the most difficult, hard-won but necessary decision in his life.} and in Germany one had read: {\it German science appeals to the intelligentsia of the whole world to cede their understanding to the stiving German nation - united by Adolf Hitler - for freedom, honor, justice and peace, to the same extent as they would for their own nation}.
 
All these circumstances can be recognized in what Ising had to overcome until his rescue. I believed that his fate would never happen anymore to people in Europe.


\begin{appendix} 
\section{Reviews concerning Ising and his model}

In this appendix time line of the literature concerning the Ising model and Ising's life is given.
The references contained in there leads to further and deeper connections of the  relevant historical and scientific circumstances.

\begin{enumerate}
\item[1968] Brush, {\it  History of the Lenz-Ising Model.} Reviews of Modern Physics, {\bf 39} 883 -  893 
\item[1983] Brush, {\it Statistical Physics and the Atomic Theory of Matter from Boyle and Newton to Landau and Onsager.} Princeton University Press
\item[1986] J. Ising, {\it Walk on a Tightrope or Paradise Lasted a Year and a Half} unpublished manuscript \url{http://www.icmp.lviv.ua/ising/books.html}
\item[1992] Ed. L. Hoddeson, E. Braun, J. Teichmann and Sp. Weart {\it Out of the Crystal Maze} Oxford University Press New York Oxford 
\item[1992] Keh - Ying Lin, {\it Spontaneous Magnetization of the Ising Model} Chinese Journal of Physics {\bf 30}/3, 287 - 319 
\item[1995] S. M. Bhattacharjee and A. Khare {\it Fifty Years of the Exact Solution of the Two-dimensional Ising Model by Onsager}  Curr.Sci. {\bf 69}   816 - 820
\item[1995] S. Kobe, {\it Ernst Ising zum 95. Geburtstag} Phys. Bl. {\bf 51} 426 
\item[1995] L. M. Robinson, {\it The Ising Model: Its Creator, Development, and Application} manuscript; Ernst Ising Collection at Bradley University Library (Peoria, Ill.).
\item[1996] C. Domb. {\it The Critical Point A historical introduction to the modern theory of critical phenomena} Taylor \& Francis, London
\item[1996] Martin H. Krieger, {\it Constitutions of Matter: Mathematically Modeling the Most Everyday of Physical Phenomena} University of Chicago Press Chicago and London
\item[1997] S. Kobe, {\it Ernst Ising - Physicist and Teacher} J. Stat. Phys. {\bf 88} 991 - 995
\item[1998] S. Kobe, {\it Das Ising-Modell - gestern und heute} Phys. Bl. {\bf 54} 917 - 920
\item[1999] R. I. G. Hughes, {\it The Ising model, computer simulations, and universal physics} in 
Morgan and Morrison (Eds.) Models as Mediators. Cambridge University Press,
\item[2000] S. Kobe, {\it Ernst Ising 1900 - 1998} Brazil. J. Phys. {\bf 30} 649 - 653
\item[2005] M. Niss, {\it Phenomena, Models and Understanding. The Lenz - Ising Model and Critical Phenomena 1920-1971.} Thesis Department of Mathematics and Physics (IMFUFA) at the University of Roskilde, Denmark
\item[2005] M. Niss, {\it History of the Lenz - Ising Model 1920 - 1950:
From Ferromagnetic to Cooperative Phenomena} Arch. Hist. Exact Sci. {\bf 59} 267 - 318
\item[2009] M. Niss, {\it History of the Lenz - Ising Model 1950-1965:
from irrelevance to relevance} Arch. Hist. Exact Sci. {\bf 63} 243 - 287
\item[2011] M. Niss, {\it History of the Lenz - Ising model 1965 - 1971: the role of a simple model in understanding critical phenomena} Arch. Hist. Exact Sci. {\bf 65} 625 - 658
\item[2017] T. Ising, R. Folk, R. Kenna, B. Berche, Yu. Holovatch, {\it The Fate of Ernst Ising and the Fate of His Model} J. Phys. Studies {\bf 21} 3002- 19p.
\end{enumerate}

\end{appendix}

\bibliographystyle{ws-rv-van}


\end{document}